\documentclass[12pt]{article}

\usepackage{amsmath,amssymb,graphicx} 
\usepackage{epsf}
\usepackage{pstricks}
\usepackage{axodraw}
\usepackage{cite}

\newcommand{\BL}{}

\newcommand{\beq}{\begin{eqnarray}}
\newcommand{\eeq}{\end{eqnarray}}

\newcommand{\centeron}[2]{{\setbox0=\hbox{#1}\setbox1=\hbox{#2}\ifdim
                           \wd1>\wd0\kern.5\wd1\kern-.5\wd0\fi \copy0
                           \kern-.5\wd0\kern-.5\wd1\copy1\ifdim\wd0>\wd1
                           \kern.5\wd0\kern-.5\wd1\fi}}
\newcommand{\ltap}{\>\centeron{\raise.35ex\hbox{$<$}}
                   {\lower.65ex\hbox{$\sim$}}\>}
\newcommand{\gtap}{\>\centeron{\raise.35ex\hbox{$>$}}
                   {\lower.65ex\hbox{$\sim$}}\>}

\newcommand\ZZ{\hbox{\zfont Z\kern-.4emZ}}
\font\zfont = cmss10 
\newcommand{\sfrac}[2]{{\textstyle\frac{#1}{#2}}}

\def\tv#1{\vrule height #1pt depth 5pt width 0pt}

\newcommand{\drawsquare}[2]{\hbox{%
\rule{#2pt}{#1pt}\hskip-#2pt
\rule{#1pt}{#2pt}\hskip-#1pt
\rule[#1pt]{#1pt}{#2pt}}\rule[#1pt]{#2pt}{#2pt}\hskip-#2pt
\rule{#2pt}{#1pt}}

\newcommand{\Yfund}{\raisebox{-.5pt}{\drawsquare{6.5}{0.4}}}
\newcommand{\Ysymm}{\raisebox{-.5pt}{\drawsquare{6.5}{0.4}}\hskip-0.4pt%
        \raisebox{-.5pt}{\drawsquare{6.5}{0.4}}}

\begin{document}
\begin{titlepage}
\begin{flushright}
{\tt hep-ph/0305237} \\
Saclay t03/061\\
\end{flushright}

\vskip.5cm
\begin{center}
{\huge \bf Gauge Theories on an Interval:}
\vskip.1cm
{\huge \bf  Unitarity without a Higgs}
\vskip.1cm
\vskip.2cm
\end{center}
\vskip0.2cm

\begin{center}
{\bf
{Csaba Cs\'aki}$^{a}$,
{Christophe Grojean}$^{b}$, {Hitoshi Murayama}$^{c}$,
\\
{Luigi Pilo}$^{b}$
{\rm and}
John Terning$^{d}$}
\end{center}
\vskip 8pt

\begin{center}
$^{a}$ {\it Newman Laboratory of Elementary Particle Physics\\
Cornell University, Ithaca, NY 14853, USA } \\
\vspace*{0.1cm}
$^{b}$ {\it Service de Physique Th\'eorique,
CEA Saclay, F91191 Gif--sur--Yvette, France} \\
\vspace*{0.1cm}
$^{c}$ {\it Department of Physics, University of California at Berkeley and \\
Lawrence Berkeley National Laboratory, Berkeley, CA 94720, USA} \\
\vspace*{0.1cm}
$^{d}$ {\it Theory Division T-8, Los Alamos National Laboratory, Los Alamos,
NM 87545, USA} \\
\vspace*{0.3cm}
{\tt  csaki@mail.lns.cornell.edu,  grojean@spht.saclay.cea.fr,
murayama@lbl.gov, pilo@spht.saclay.cea.fr, terning@lanl.gov}
\end{center}

\vglue 0.3truecm

\begin{abstract}
\vskip 3pt
\noindent We consider extra dimensional gauge theories on an interval.
We first review the derivation of the consistent
boundary conditions (BC's) from the action principle. These
BC's include choices
that give rise to breaking of the gauge symmetries. The
boundary conditions could be chosen to coincide with those
commonly applied in orbifold theories, but there are many
more possibilities.
To investigate the nature
of gauge symmetry breaking via BC's we calculate
the elastic scattering amplitudes for longitudinal gauge bosons. We find that
using a consistent set of BC's
the terms in these amplitudes that explicitly
grow with energy always cancel without
having to introduce any additional scalar degree of freedom, but rather by the
exchange of Kaluza--Klein (KK) gauge bosons. This suggests that perhaps
the SM Higgs could be completely eliminated in favor of some KK towers of gauge
fields. We show that from the low-energy effective theory perspective this
seems to be indeed possible. We display an extra dimensional toy model,
where BC's introduce a symmetry breaking pattern and mass
spectrum that resembles that in the standard model.

\end{abstract}

\end{titlepage}

\newpage


\section{Introduction}
\label{sec:intro}
\setcounter{equation}{0}
\setcounter{footnote}{0}

A crucial ingredient of the standard model of particle physics is the
Higgs scalar.
One of the main arguments for the existence of the Higgs
is~\cite{LlewellynSmith, Dicus, GBequivalence, SMunitarity, Chanowitz}
that without it the scattering amplitude for the longitudinal
components of the massive $W$ and $Z$ bosons would grow with energy as
$\sim E^2$, and thus violate unitarity at energies of order
$4\pi M_W/g \sim 1.5$~TeV. It has been shown in~\cite{Sekhar1,Sekhar3} that
higher dimensional gauge theories
maintain
unitarity in the sense that the terms in the
amplitude that would grow
with energies as $E^4$ or $E^2$ cancel (though the theory
itself becomes strongly interacting at a cutoff scale which depends on the
size of the extra dimension and the effective gauge coupling,
and usually tree-level unitarity also breaks down at a scale related to the
cutoff scale due to the growing number of KK modes that can contribute to the
constant pieces of certain amplitudes).
For a related discussion see~\cite{DeCurtis}. This on its own is not
so surprising, since one would naively expect that higher dimensional
gauge theories behave well in the energy range where they can be valid
effective theories. However, such higher dimensional theories can also be used
to break the gauge symmetries if one compactifies the theory on an
interval instead of a circle. Then by assigning non-trivial boundary conditions (BC's)
to the gauge fields at the endpoints of the interval one can reduce the
number of unbroken gauge symmetries, and thus effectively generate gauge boson
masses even for the modes that would remain massless when only
Neumann BC's are imposed. This then raises the question,
whether the cancellation in the scattering amplitudes of the terms that
grow with energy is maintained or not in the presence of such breaking
of gauge symmetries. This issue is related to the question of whether
the breaking of gauge invariance via boundary conditions is soft or hard.
We would like  to give a general analysis of this question (which also have
been recently addressed
in some particular examples in Refs.~\cite{hitoshi,haba},
see also~\cite{Schwartz} for a related discussion in the case of KK gravity).

In this paper we investigate the nature of gauge symmetry breaking via
BC's. First we review
the derivation of  the set of equations that the boundary
conditions have to obey in order to minimize the action, including a
discussion of the issue of gauge fixing.
The possible set of BC's (BC's) include the commonly
considered orbifold\footnote{In
orbifold theories one starts with a
theory on a circle and then projects out some states not invariant under a
symmetry of the theory on the circle.} BC's~\cite{orbifold,Kawamura,HebeckerJMR,Quiros}
but as it was already noted in~\cite{HebeckerJMR} there are more possibilities.
 For example,
it is easy to reduce the rank of the gauge group with more general
BC's~\cite{HebeckerJMR}.
The question that such theories raise is whether such a breaking
of the gauge symmetries via BC's yields a consistent theory or not.
Since we are insisting that the BC's be consistent with the
variation of a gauge invariant Lagrangian that has no explicit gauge symmetry
breaking, one would guess that such breaking should be soft. In order
to verify this, we investigate in detail the issue
of unitarity of scattering amplitudes in such 5D gauge theories compactified on
an interval, with non-trivial BC's. We derive the general
expression for the amplitude for elastic scattering of longitudinal
gauge bosons, and write down the necessary
conditions for the cancellation of the terms that grow with energy.
We find that all the consistent
BC's  are unitary in the sense that all terms proportional to $E^4$ and
$E^2$ vanish.
In fact, any theory with only Dirichlet or Neumann BC's
is unitary. Surprisingly, this would also include theories where the boundary
conditions can be thought of as coming from a very large expectation value
of a brane localized Higgs field, in the limit when the expectation value
diverges. For such theories with ``mixed" BC's, even when the
$E^4$ terms cancel, the $E^2$ term in
the amplitude does not cancel in general. This is not surprising, since such
mixed BC's would generically come from an explicit mass term for the gauge
field localized on the boundary. Thus cancellation of the $E^2$ term would
happen only, if the explicit mass term is completed into a gauge invariant
scalar mass term, in which case the exchange of
these boundary scalar fields themselves have to be
included in order to recover a good high-energy behavior for the theory.
Indeed, we find that in some cases it may
be possible to introduce such boundary degrees of freedom which would exactly
enforce the given BC's, and their contribution cancels
the remaining amplitudes that grow with energy.

These arguments suggest that it should be possible to build an effective theory
which has no Higgs field present at all, but where the unitarity
of gauge boson scattering amplitudes is ensured by the presence of
additional massive gauge fields. We show a simple example of an effective
theory of this sort where a single KK mode for the $W$'s and the $Z$ is
needed to ensure unitarity, and which are sufficiently heavy and
sufficiently weakly coupled to have evaded direct detection and would not
have contributed much to electroweak precision observables.

In order to actually make an effective theory of this sort appealing one
would have to give a UV completion, for example  at least
in  terms of an extra dimensional theory. Therefore, we will consider
several toy models of symmetry breaking with extra dimensions. The first two
models are prototype examples of orbifold vs. brane Higgs breaking of
symmetries, which we combine in the final semi-realistic model based on
the breaking of a left-right symmetric model by an orbifold using an outer
automorphism. This model is similar to the standard
model, in that it has unbroken electromagnetism, the lightest massive gauge
bosons resemble the $W$ and the $Z$, and their mass ratio could be
close to a realistic one. However, the masses of the KK modes are too light,
and their couplings too strong. Nevertheless, we view this model as a step
toward a realistic theory of electroweak symmetry breaking without a Higgs
boson.

\section{Gauge Theories on an Interval: Consistent BC's}
        \label{sec:general}
\setcounter{equation}{0}
\setcounter{footnote}{0}

We consider a theory with a single extra dimension compactified on
an interval with endpoints $0$ and $\pi R$. We want to study in
this section what are the possible BC's that the bulk gauge fields
have to satisfy. We denote the bulk gauge fields by  $A_M^a(x,y)$,
where $a$ is the gauge index, $M$ is the Lorentz index
$0,1,2,3,5$, $x$ is the coordinate of the ordinary four dimensions
and $y$ is the coordinate along the extra dimension (we will use
from time to time a prime to denote a derivative with respect to
the $y$ coordinate). We will assume a flat space-time background.
We will consider several cases in this section. First we will look
at the simplest example of a scalar field in the bulk, which does
not have any of the complications of gauge invariance and gauge
fixing. Then we will look at pure 5D Yang-Mills theory on an
interval. The cases of gauge theory with a bulk or brane scalar
field are discussed in Appendix A.

\subsection{Bulk scalar}
To start out, let us consider a bulk scalar $\phi$ field
on an interval with an action
\begin{equation}
\mathcal{S}= \int d^4x\,\int_0^{\pi R} dy \left( \frac{1}{2} \partial^{{}_N} \phi
\partial_{{}^N} \phi - V(\phi) \right) +  \int_{y=0} d^4x \, \frac{1}{2} \phi^2 \, M_1^2 +
\int_{y= \pi R} d^4x \, \frac{1}{2} \phi^2 \, M_2^2 \, .
\label{bsc}
\end{equation}
In  order to find the consistent set of BC's we impose that the variation of (\ref{bsc})
vanishes. Varying the action and integrating by parts
we get
\begin{eqnarray}
\delta \mathcal{S}  = && - \int d^4x\,\int_0^{\pi R} dy \; \delta \phi \,
\left( \Box_5
\phi +\frac{\partial V}{\partial \phi}\right) - \nonumber \\
&&
\int d^4x \left( \delta \phi \left(\partial_5 \phi + M^2_1 \, \phi \right)
\big |_{\pi R} -  \delta \phi \left(\partial_5 \phi + M^2_2 \, \phi \right)
\big |_0 \right)=0 \, .
\end{eqnarray}
Note, that we kept the boundary terms obtained from integrating
by parts along the finite extra dimension, while we have assumed
as usual that the fields and their derivatives die off as $x_\mu \to \infty$.
The variation of the bulk terms will give the usual bulk equation of motion
\begin{equation}
\Box_5
\phi +\frac{\partial V}{\partial \phi}=0.
\end{equation}
In order to ensure that the action is minimized, one also has to ensure that
the variation of the action from the boundary pieces also vanish:
\begin{equation}
\delta \phi \left(\partial_5 \phi + M^2_i \, \phi \right)
\big |_{0,\pi R} =0.
\end{equation}
A consistent BC is one that automatically enforces the
above equation. There are two ways to solve this equation: either the
variation of the field on the boundaries is zero, or the expression
multiplying the variation.
Therefore the
 consistent set of BC's that respects 4D Lorentz invariance at
$y=0,\pi R$ are
\begin{eqnarray}
        \label{SM}
&{\rm ({\it i})}   & \left(\partial_5 \phi + M^2_i \, \phi \right)|_{y=0,\pi R}=0\, ; \\
\label{SD}
&{\rm ({\it ii})}   & \phi|_{y=0,\pi R} = const.\, .
\end{eqnarray}
Eq.~(\ref{SM}) corresponds to a mixed BC and it reduces to
Neumann for vanishing
boundary mass $M^2_i$; the value of $\phi$ at $y=0,\pi R$ is not specified.
The second type of BC corresponds to fixing the values of $\phi$ on the boundary,
and reduces to a Dirichlet BC when $const.=0$.
It is a matter of choice which one of these conditions one is imposing
on the boundaries. One could pick one of these conditions at $y=0$
and the other at $y=\pi R$. When several scalar fields are present, there is also the more interesting
possibility to cancel the sum of all the boundary terms without having to require
that individually each term is vanishing by itself. We will see an example for this in Section~\ref{sec:LR}
of this paper.

\subsection{Pure gauge theory in the bulk}
In this case the 5D gauge boson will decompose into a 4D gauge
boson $A_\mu^a$ and a 4D scalar $A_5^a$ in the adjoint representation.
Since there is a quadratic term mixing $A_\mu$ and $A_5$
we need to add a gauge fixing term that eliminates this cross term.
Thus we write the action after gauge fixing in R$_\xi$
gauge as
\begin{equation}
        \label{eq:lagr}
\mathcal{S}=
\int d^4x\,\int_0^{\pi R} dy \left(
-\frac{1}{4} F_{\mu\nu}^a F^{a\mu\nu} -\frac{1}{2} F_{5\nu}^a F^{a5\nu}-
\frac{1}{2\xi} (\partial_\mu A^{a\mu} -\xi \partial_5 A_5^a)^2 \right),
\end{equation}
where $F^a_{MN}=\partial_M A_N^a -  \partial_N A_M^a
+g_5 f^{abc}\, A^b_M A^c_N$, and the $f^{abc}$'s
are the structure constants of the gauge group.
The gauge fixing term is chosen such that (as usual) the cross terms
between the 4D gauge fields $A_\mu^a$ and the 4D scalars $A_5^a$ cancel
(see also \cite{Muck}).
Taking $\xi\to \infty$ will result in the unitary gauge, where all the KK modes
of the scalars fields $A^a_5$ are unphysical (they become the longitudinal
modes of the 4D gauge bosons), except if there is a zero mode for the
$A_5$'s. We will assume that every $A_5^a$ mode is massive, and thus that all
the $A_5$'s are eliminated in unitary gauge.

The variation of the action~(\ref{eq:lagr})  leads, as usual after integration by parts, to
the bulk equations of motion as well as to boundary terms (we denote by $[F]$ the boundary quantity
$F(\pi R)- F(0)$):
\begin{eqnarray}
\delta \mathcal{S} & = &
\int d^4x\, dy \left(
\partial_M F^{a M\nu} - g_5 f^{abc}\, F^{bM\nu}A_M^c
+ \frac{1}{\xi} \partial^\nu \partial^\sigma A_\sigma^a
- \partial^\nu \partial_5 A_5^a  \right) \delta A_\nu^a
\nonumber\\
&&
- \int d^4x\, dy \left(
\partial^\sigma F^{a}_{\sigma 5} - g_5 f^{abc}\, F^{b}_{\sigma 5} A^{c \sigma}
+\partial_5 \partial_\sigma A^{a \sigma} - \xi \partial^2_5 A^a_5 \right) \delta A^a_5
\nonumber\\
&&
+\int d^4x \left( \left[ F^{a}_{5 \nu} \, \delta A^{a \nu} \right]
+ \left[ (\partial_\sigma A^{a\sigma} - \xi \partial_5 A^a_5) \delta A^a_5 \right] \right).
\end{eqnarray}
The bulk terms will give rise to the usual bulk equations of motion:
\begin{eqnarray}
\label{bulkeom}
&&\partial_M F^{a M\nu} - g_5 f^{abc}\, F^{bM\nu}A_M^c
+ \frac{1}{\xi} \partial^\nu \partial^\sigma A_\sigma^a
- \partial^\nu \partial_5 A_5^a =0,
\nonumber \\
&&\partial^\sigma F^{a}_{\sigma 5} - g_5 f^{abc}\, F^{b}_{\sigma 5}
A^{c \sigma}
+\partial_5 \partial_\sigma A^{a \sigma} - \xi \partial^2_5 A^a_5=0 .
\end{eqnarray}
However, one has to ensure that the variation of the boundary pieces
vanish as well. This will lead to the requirements
\begin{eqnarray}
&&  F^{a}_{\nu 5} \, \delta A^{a \nu}{}_{|0,\pi R} =0,
\\
        \label{bc}
&&  (\partial_\sigma A^{a\sigma} - \xi \partial_5 A^a_5) \delta A^a_5{}_{|0,\pi R} =0.
\end{eqnarray}
The BC's have to be such that the above equations be satisfied.
For example, one can fix all the variations of the fields to vanish
at the endpoints, in that case the above boundary terms are clearly
vanishing. However, one can also instead of setting
$ \delta A^a_5{}_{|0,\pi R} =0$ require that its coefficient
$\partial_\sigma A^{a\sigma} - \xi \partial_5 A^a_5$ vanishes. The
different choices lead to different consistent BC's
for the gauge theory on an interval.
There are different generic choices of BC that ensure the
cancellation of variation at the boundary and preserve 4D Lorentz invariance:
\begin{eqnarray}
        \label{eq:DD}
&{\rm ({\it i})}   & A^a_{\mu |} = 0, A^a_{5 |} =const.\ ; \\
        \label{eq:DN}
&{\rm ({\it ii})}  & A^a_{\mu |} = 0, \partial_5 A^a_{5 |} =0 \ ;\\
        \label{eq:ND}
&{\rm ({\it iii})} & \partial_5 A^a_{\mu |} =0, A^a_{5 |}= const. \ .
\end{eqnarray}
However, besides these general choices where the individual terms in the sum
of (\ref{bc}) vanish, there could be more interesting situations where
only the sum of the terms in (\ref{bc}) vanish. We will see an example for
this in Section 6 of this paper. While conditions (\ref{eq:DD}) and (\ref{eq:DN}) are
 exact,
condition (\ref{eq:ND}) only satisfies Eq. (\ref{bc}) to linear order. The exact
solution
requires $F^{a}_{\nu 5} =0$ which imposes additional constraints especially in
the case of an unbroken gauge symmetry (i.e. zero mode gauge fields).

Generically, it is a choice which of these BC's one wants
to impose. One can impose a different type of condition
for  every different field, meaning for the different colors $a$ and
for $A_\mu$ vs. $A_5$, as long as certain consistency conditions
related to the gauge invariance of massless gauge fields is obeyed.
Different choices correspond to different physical situations. An
analogy for this is the theory of a vibrating rod. The equation of
motion is obtained from a variational principle minimizing the total
mechanical energy, including the terms appearing from the boundaries.
Which of the BC's one  chooses depends on the
physical circumstances at its ends: if it is fixed at both ends
the BC is clearly that the displacement at the endpoints is
zero. However, if it is fixed only on one end, then one will get a non-trivial
equation that the displacement on the boundaries has to satisfy, the analog of
which is the condition $F^a_{\nu5}=$ in (\ref{bc}).

Similarly in the case of gauge theories, it is our choice what
kind of physics we are prescribing at the boundaries, as long as
the variations of the boundary terms vanish. A priori, one can
impose different BC's for different gauge directions. However,
some consistency conditions on the gauge structure may exist. For
instance, if one wants to keep a massless vector, then in order to
preserve 4D Lorentz invariance the action should possess a gauge
invariance. This means that the massless gauge bosons should form
a subgroup of the 5D gauge group.

One should note that there is a wider web of consistent BC's
than the one encountered in orbifold theories. For instance, within each gauge direction,
the BC~(\ref{eq:DD}) would have never been consistent
with the reflection symmetry $y\to -y$ symmetry of an orbifold.
The full gauge structure of the BC's is also much less
constrained: in orbifold theories, the gauge structure was dictated by the use
of an automorphism of the Lie algebra, which was a serious obstacle
in achieving a rank reducing symmetry breaking. As we will see explicitly in
Sections~\ref{sec:BraneHiggs} and~\ref{sec:LR}, these difficulties are easily alleviated when considering
the most general BC's~(\ref{eq:DD})-(\ref{eq:ND}).

In order to actually quantize the theory one also needs to add the
Faddeev--Popov ghosts to the theory. One can add 5D FP ghost fields $c^a$ and
$\bar{c}^a$ using the gauge fixing function from (\ref{eq:lagr})
\begin{equation}
{\cal L}_{FP}=\bar{c}^a\left( -(\partial_\mu D^\mu)^{ab} +\xi (\partial_5 D_5)
\right) c^b
\end{equation}
The ghost fields have their own boundary conditions as well.

The cases for gauge theories with bulk scalars or localized scalars are
discussed in Appendix A. For the case of a scalar localized at the endpoint
the generic form of the BC for
the gauge fields (in unitary gauge) will be of the form
\begin{equation}
\partial_5 A_\mu^a{}_{|0,\pi R}= V_{0,\pi R}^{ab} A^b_\mu {}_{|0,\pi R}.
\label{mixedbc}
\end{equation}
These are mixed BC's that still ensure the hermiticity (self-adjointness)
of the Hamiltonian. In the limit $V^{ab} \rightarrow 0$ the mixed BC reduces to
a Neumann BC, while the limit $V^{ab} \rightarrow \infty$ produces a Dirichlet BC.

Finding the KK decomposition of the gauge field reduces to solving a Sturm--Liouville problem
with Neumann or Dirichlet BC's, or in the case of boundary
scalars with mixed BC's.
Those general BC's lead to a Kaluza--Klein expansion of the gauge fields of
the form
\begin{equation}
A_\mu^a(x,y)=\sum_n \epsilon_\mu\, f_n^a(y) e^{ip_n x} ,
\end{equation}
where $p_n^2=M_n^2$ and $\epsilon_\mu$ is a polarization vector.
These wavefunctions (due to the assumption of 5D Lorentz invariance, {\it i.e.}, of a flat
background) then satisfy the equation:
\begin{equation}
{f^a_n}''(y)+M_n^{a\, 2} f^a_n (y)=0, \ \  {f^a_n}'(0,\pi R) =V_{0,\pi R}^{ab} f^b_n (0,\pi R).
\end{equation}
The couplings between the different KK modes can then be
obtained by substituting this expression into the Lagrangian (\ref{eq:lagr})
and integrating over the extra dimension. The resulting couplings are then
the usual 4D Yang-Mills couplings, with the gauge coupling $g_4$ in the
cubic and gauge coupling square in the quartic vertices replaced by the
effective couplings involving the integrals of the wave functions
of the KK modes over the extra dimension:
\begin{eqnarray}
        \label{eq:cubic}
g_{cubic} & \to & g^{abc}_{mnk}=g_5 \int dy f_m^a (y) f_n^b (y) f_k^c (y),
\\
        \label{eq:quartic}
g_{quartic}^2 & \to & g^{2\, abcd}_{mnkl}=g_5^2  \int dy f_m^a (y) f_n^b (y) f_k^c (y) f_l^d (y).
\end{eqnarray}
Here $a,b,c,d$ refer to the gauge index of the gauge bosons and $m,n,k,l$ to
the KK number.

\section{Unitarity of the Elastic Scattering Amplitudes for Longitudinal
Gauge Bosons}
        \label{sec:amplitude}
\setcounter{equation}{0}
\setcounter{footnote}{0}

We have seen above that gauge theories with BC's lead to various
patterns of gauge symmetry breaking. The obvious question is whether these
should be considered spontaneous (soft) or explicit (hard) breaking of the
gauge invariance. Since we have obtained the BC's from varying a gauge
invariant Lagrangian, one would expect that the breaking should be soft.
We will now investigate this question by examining the high-energy behavior
of the elastic scattering amplitude of longitudinal
gauge bosons in the theory described in the previous section.
Using the couplings obtained above we are now ready to
analyze these amplitudes.
First we will extract the terms that grow with energy in these amplitudes,
then discuss what BC's will enable us to cancel the terms
which grow with energies. We will restrict our analysis to elastic scattering.\footnote{
Working in the unitary gauge, there is a pole in the inelastic scattering amplitude
when a massless gauge boson is exchanged in the $u$- and $t$-channel. Technically, this requires
to work in a general $\xi$ gauge and more computations are needed to derive
sum rules equivalent to the ones we present for the elastic scattering.}

\subsection{The elastic scattering amplitude}

\begin{figure}[ht]
\centerline{\includegraphics[width=0.4\hsize]{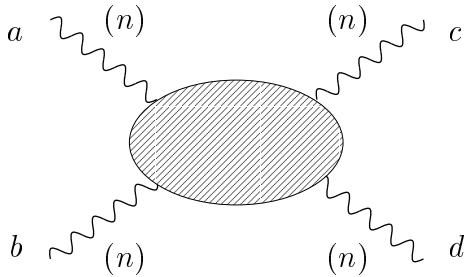}}
\caption[]{Elastic scattering of longitudinal modes of KK gauge bosons,  $n+n\to n+n$,
with the gauge index structure $a+b\to c+d$.}
\label{fig:scattering}
\end{figure}

\begin{figure}[!ht]
\centerline{\includegraphics[width=0.75\hsize]{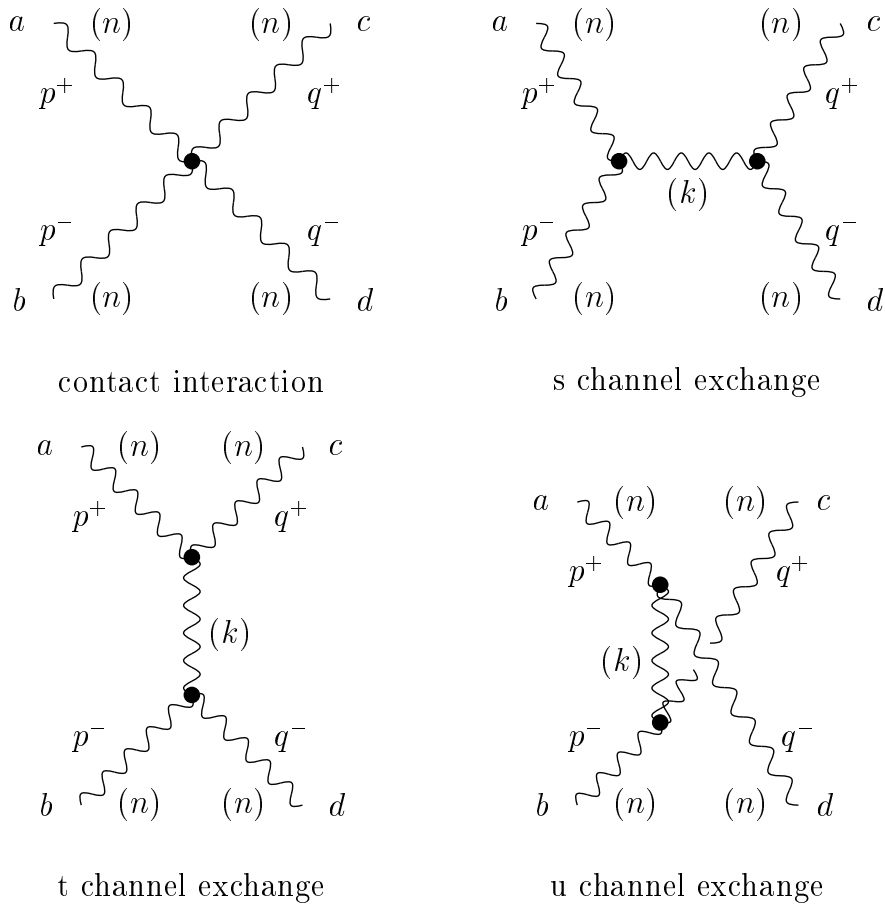}}
\label{fig:diagrams}
\caption{The four gauge diagrams  contributing at tree level
to the gauge boson elastic scattering amplitude.}
\end{figure}

We want to calculate the energy dependence
of the amplitude of the elastic scattering of the longitudinal modes of the
KK gauge bosons $n+n\to n+n$ with gauge index structure $a+b\to c+d$ (see Fig.~\ref{fig:scattering}),
where this
process involves both exchange of the $k$'th KK mode from the cubic vertex,
and the direct contribution from the quartic vertex. We will also assume,
that the BC's corresponding to the external modes with gauge indices
$a,b,c,d$ are of the same type, that is they have the same KK towers (however
we will {\it not} assume this for the modes that are being exchanged).
There are four
diagrams as shown in Fig.~\ref{fig:diagrams}: the $s$, $t$ and $u$-channel exchange
of the KK modes, and the contribution of the quartic vertex.
The kinematics assumed for this elastic scattering is in the center of mass
frame, where the incoming momentum vectors are $p_\mu=(E,0,0,\pm \sqrt{E^2-M_n^2})$,
while the outgoing momenta are $(E,\pm\sqrt{E^2-M_n^2} \sin \theta,0,
\pm\sqrt{E^2-M_n^2} \cos \theta )$. $E$ is the incoming energy, and $\theta$
the scattering angle with forward scattering for $\theta =0$.
The longitudinal polarization vectors are as usual
$\epsilon_\mu =(\frac{|\vec{p}|}{M},\frac{E}{M} \frac{\vec{p}}{|\vec{p}|})$
and accordingly the contribution of each diagram can be as bad as
$E^4/M_n^4$. It is straightforward to evaluate the full scattering amplitude, and
extract the leading behavior for large values energies of this amplitude. The general structure of the
expansion in energy contains three terms:
 \begin{equation}
\mathcal{A}= A^{(4)} \frac{E^4}{M_n^4} +
A^{(2)} \frac{E^2}{M_n^2}+
A^{(0)}+{\cal O}\left(\frac{M_n^2}{E^2}\right).
 \end{equation}
%

It may seem inconsistent to formally expand the amplitude in energies, when
for any fixed energy there are KK modes that are much heavier and the series is
potentially non-convergent.  However from the higher dimensional effective
field theory we can
see that the heaviest modes are not important. Summing over all the modes is
just the simplest way of maintaining gauge invariance which would be broken by
a hard cutoff on the spectrum.

To show this consider a non-Abelian gauge theory in $D$ dimensions
with a cutoff $\Lambda$. The leading effect of integrating out the
KK modes above $\Lambda$ should be given by gauge invariant higher
dimensional operators. An example of such operators is given by
\beq \label{higherdim} \frac{g_D^3}{\Lambda^{6-D}} F_{MN}^3~. \eeq
All other gauge invariant operators of the same dimension should
give similar results for the scattering amplitudes. The
coefficient of this operator can be fixed by first going into the
normalization where the coefficient of the kinetic term is $-
1/(4g_D^2)$. In this normalization $F_{MN}$ has dimension two,
thus the prefactor $1/\Lambda^{6-D}$. Going back to canonical
normalization we get the higher order term of (\ref{higherdim}).
Alternatively, we can see that this operator contains three gauge
fields, so if it comes from a loop of massive gauge fields it
should contain three gauge couplings. The contribution to
longitudinal scattering from the ordinary $F_{MN}^2$ term is
proportional to \beq g_D^2 \left(E R\right)^4 \eeq (where we have
assumed the incoming gauge boson mass is ${\cal O}(1/R)$,  the
inverse of the compactification radius). This can be seen easiest
by looking at the four-point coupling of the gauge field. That has
an explicit factor of $g_D^2$, and there are four polarization
vectors which are of order $E/M\sim ER$. Naively, the contribution
from the higher dimensional operator is growing faster with
energy, with a power $E^6$. However, gauge invariance will
actually soften this amplitude. One can see this explicitly by
noting that from (\ref{higherdim}) one needs two factors of
$\partial_\mu A_\nu -\partial_\nu A_\mu$, and two other factors of
the gauge field. This will imply that two of the polarization
vectors appear in the combination $p_\mu \epsilon_\nu -p_\nu
\epsilon_\mu$, which after explicitly substituting for the
polarization vectors is just proportional to the mass of the
external particle, rather than growing with energy. Therefore the
contribution from the $F_{MN}^3$ terms scales as \beq
\frac{g_D^4}{\Lambda^{6-D}} \left(E R \right)^2 \frac{1}{R^2}\, ,
\eeq so the ratio of the correction term to the leading term is
\beq \frac{g_4^2}{(\Lambda R)^{6-D} (ER)^2}\, , \eeq where we have
used $g_D^2=g_4^2 R^{D-4}$.  Thus for $E \gg 1/R$, $\Lambda R> 1$,
and $D<6$ the contributions from the highest KK modes are
suppressed. Note that here we have not used the non-trivial
cancellation of the $E^4$ term in the amplitude. Once that is
taken into account, the effects of the KK modes are still
suppressed as long as $D<6$ and $\Lambda R > 1$. For a 5D theory
$\Lambda R$ can be large as $24 \pi^3$ \cite{Chacko:1999mi}.

We would like to understand under what circumstances $A^{(4)}$ and
$A^{(2)}$ vanish. The vanishing of these coefficients would ensure
that there are no terms in the amplitude that explicitly grow with
energy. This is a necessary condition for the tree-level unitarity
of a theory. However, this is certainly not a sufficient
condition. The finite piece in the amplitude $A^{(0)}$ should also
not be too large in order for the theory to be tree-unitary, for
every possible scattering. In theories with extra dimensions,
since there are an infinite number of KK modes available, there
will always be some amplitudes that have finite pieces, which
however grow as the number of exchanged KK modes is allowed to
increase. This simply reflects the higher dimensional nature of
the theory, and the fact that higher dimensional gauge theories
are non-renormalizable. Therefore, tree unitarity is expected to
break down at energies comparable to the 5D cutoff scale, even
when $A^{(4)}=A^{(2)}=0.$ When discussing unitarity of these
models we will simply mean the cancellation of the $A^{(4)}$ and
$A^{(2)}$ amplitudes. We stress again that this does not imply
that the conventional unitarity bounds on the finite amplitudes
have to be satisfied for all processes.

The largest term, growing with $E^4$, depends only on the effective couplings
and not on the mass spectrum:
\begin{eqnarray}
A^{(4)}= i \left(  g^2_{nnnn} - \sum_k  g_{nnk}^2  \right)
\left( f^{abe} f^{cde}  (3+6 \cos \theta - \cos^2 \theta)
 +2   (3- \cos^2 \theta)
 f^{ace} f^{bde}          \right).
\label{E4}
\end{eqnarray}
The expression for the amplitude  that grows with $E^2$  is:
\begin{eqnarray}
A^{(2)}=
\frac{i}{M_{n}^2} f^{ace} f^{bde}
\left(
4   g_{nnnn}^2 M_{n}^2-3\sum_k  g_{nnk}^2 M_k^2
\right)
-\frac{i}{2 M_{n}^2} f^{abe} f^{cde}
\left(
4    g_{nnnn}^2  M_{n}^2 \vphantom{-3\sum_k M_k^2 g_{nnk}^2}
\right.
\nonumber
\\
\left.
-3\sum_k  g_{nnk}^2 M_k^2
+\left(
12   g_{nnnn}^2 M_{n}^2 + \sum_k g_{nnk}^2 (3 M_k^2-16M_n^2) \right)
\cos \theta
\right).
\label{E2}
\end{eqnarray}
Both in (\ref{E4}) and (\ref{E2}) everywhere below
the KK indices $n$ and $k$ actually have to be interpreted
as  double indices, that stand both for the color index and the KK number
of the given color index, {\it e.g.}, $k \sim (k,e)$. Also, as stated above,
we are assuming that
all the in-going and out-going gauge bosons satisfy the same boundary
condition and thus the
$n$ index is color-blind. In getting the expressions~(\ref{E2}) or (\ref{E2simple}) for $A^{(2)}$,
we used the Jacobi identity which requires a sum over the gauge index of the exchanged gauge boson.
Thus we have implicitly assumed that the sums like
$\sum_{k} g_{nnk}^{abe} g_{nnk}^{cde}$ or  $\sum_{k} g_{nnk}^{abe} g_{nnk}^{cde} M_k^{e\,2}$
are independent of the gauge index $e$, which indeed is a true statement as we
 will show later on.

If one assumes the cancellation of the $E^4$ terms (which we will show
is indeed always the case) one has the relation
\begin{equation}
g^2_{nnnn} = \sum_k  g_{nnk}^2.
\end{equation}
Using this relation the expression for the $E^2$ terms
can be  simplified to the following:
\begin{eqnarray}
        \label{E2simple}
A^{(2)}=
\frac{i}{M_{n}^2}
\left(
4    g_{nnnn} M_{n}^2
-3 \sum_k  g_{nnk}^2     M_{k}^2
\right)
\left(
f^{ace} f^{bde}  - \sin^2 \sfrac{\theta}{2} \ f^{abe} f^{cde}
\right) .
\end{eqnarray}

Even though we will not consider constraints coming from the
finite ($E^0$) terms in the elastic scattering amplitude, for
completeness we present the relevant expression here (making use
of the conditions for cancelling the $A^{(4)}$ and $A^{(2)}$
terms):
\begin{equation}
A^{(0)}= i \sum_k g_{nnk}^2 \left(  f^{abe} f^{cde} F(M_k/M_n,\theta)
+  f^{ace} f^{bde} G(M_k/M_n,\theta) \right),
\end{equation}
with
\begin{eqnarray}
F(x,\theta) = -\frac{1}{16 \cos^2 \theta/2}
(4-10x^2+7x^4-8(4-3x^2) \cos \theta - (4-2x^2-x^4) \cos 2 \theta), \nonumber \\
G(x,\theta)= \frac{1}{2 \sin^2 \theta} (4-6x^2+3x^4+(12-10x^2+x^4) \cos^2 \theta).
\end{eqnarray}

We can summarize the above results by restating the conditions under which
the terms that grow with energy cancel:
\begin{eqnarray}
        \label{E4cancellation}
g^2_{nnnn} &=& \sum_k  g_{nnk}^2 ,\\
        \label{E2cancellation}
4   g_{nnnn}^2 M_{n}^2 &=& 3 \sum_k  g_{nnk}^2 M_{k}^2\, .
\end{eqnarray}
%

\subsection{Cancellation of the terms that grow with energy for the simplest
BC's}

The goal of the remainder of this section is to examine under what
circumstances the terms that grow with energy cancel. Consider first the
$E^4$ term. According to (\ref{E4cancellation}) the requirement for cancellation is
\begin{equation}
        \label{E4cond}
\int_0^{\pi R} dy\, f_n^4(y) = \sum_k \int_0^{\pi R} dy \int_0^{\pi R} dz\
f_n^2(y)f_n^2(z) f_k^{\vphantom{2}}(y) f_k^{\vphantom{2}}(z).
\end{equation}
One can easily see that this equation is in fact satisfied no matter what
BC one is imposing, as long as that BC
still maintains hermiticity of the differential operator $ \partial_y^2$
on $0,\pi R$. The BC maintains hermiticity of
the differential operator if it is of the form ${f'}_{|0,\pi R}= V_{0,\pi R}
f_{|0,\pi R}$. In this case one can explicitly check that $\int_0^{\pi R} h^* {g''}=
\int_0^{\pi R} {h^*}'' g$.
The reason why (\ref{E4cond}) is obeyed
is that for such hermitian operators one
is guaranteed to get an orthonormal complete set of solutions $f_k(y)$,
thus from the completeness it follows that
\begin{equation}
        \label{completeness}
\sum_k f_k(y) f_k(z) =\delta (y-z),
\end{equation}
which immediately implies (\ref{E4cond}).\footnote{Again we stress
that, restoring gauge indices, the completeness relation holds
$\sum_k f_k^e(y) f_k^e(z) =\delta (y-z)$, with no sum over $e$.
This just says that there is completeness for every gauge index
separately. This justifies the manipulations used to get
Eqs.~(\ref{E2})-(\ref{E2simple}), where the Jacobi identity was
used without paying attention to the remaining implicit gauge
indices in the sum.} The completeness relation basically implies
that any function can be expanded in terms of the eigenfunctions
of $\partial_y^2$ on the interval $0,\pi R$, $g(y) =\sum_k f_k(y)
\int_0^{\pi R} dz g(z)f_k(z)$. There is a subtlety in this if the
BC for the $f$'s is $f(0,\pi R)=0$, since in that case only
functions that are themselves zero at the boundary can be expanded
in this series. However, even in this case the expansion will
converge everywhere except at the two endpoints to the given
function, and since we will integrate over the interval anyway,
changing the function at finite number of points does not matter.
Thus we conclude that (\ref{E4cond}) always holds, the $E^4$ terms
always cancel irrespectively of the BC's imposed. Therefore, we
can now assume that the $E^4$ terms cancelled, and using the
equation that leads to the cancellation of these terms we get that
the condition for the cancellation of the $E^2$ terms is as in
(\ref{E2cancellation})
\begin{equation}
\label{E2cond}
3\, \sum_k M_k^2 \int_0^{\pi R} dy \int_0^{\pi R} dz\,
f_n^2(y)f_n^2(z) f_k^{\vphantom{2}}(y) f_k^{\vphantom{2}}(z)
=
4\, M_n^2 \int_0^{\pi R} dy \, f_n^4(y).
\end{equation}
Let us first assume that one can integrate by parts freely without picking
up any boundary terms (we will later see under what circumstances this
assumption is indeed justified). In this case we can easily derive a sum
rule of the form
\begin{equation}
        \label{eq:sumrule}
\sum_{k} M_k^{2p}
\left( \int_0^{\pi R} dy  \, f_n^2(y)  f_k^{\vphantom{2}}(y) \right)^2
= \frac{2^{2p}}{3} M_n^{2p} \int_0^{\pi R} dy f_n^4(y).
\end{equation}
Indeed, let us define
\begin{equation}
a_p^{(k)} = M_k^{2p} \int_0^{\pi R} dy\, f_n^2 (y) f_k^{\vphantom{2}}(y), \ \
b_p^{(k)} = M_k^{2p} \int_0^{\pi R} dy\, f'_{n}{}^2 (y) f_k^{\vphantom{2}}(y).
\end{equation}
Using the 5D Lorentz invariant relation between the wavefunction and the mass spectrum,
$M_k^2 f_k(y) = - f_k''(y)$, a simple integration by part gives, neglecting the boundary terms
that we will include in the next section, the recurrence relations:
\begin{equation}
a_{p+1}^{(k)} = 2 M_n^2 a_p^{(k)} -  2 b_p^{(k)}, \ \
b_{p+1}^{(k)} = 2 M_n^2 b_p^{(k)} -  2 M_n^4 a_p^{(k)},
\end{equation}
from which we obtain that
\begin{equation}
a_p^{(k)} = 2^{2p-1} M_n^{2p-2} \left(M_n^2 a_0^{(k)}- b_0^{(k)}\right).
\end{equation}
The completeness relation~(\ref{completeness}) finally allows to evaluate the sums:
\begin{equation}
\sum_k    \big(a_0^{(k)}\big)^2 = \int_0^{\pi R} dy\, f_n^4(y), \ \
\sum_k    a_0^{(k)}  b_0^{(k)}= \sfrac{1}{3} M_n^2 \int_0^{\pi R} dy\, f_n^4(y),
\end{equation}
which, combined with the recurrence relation, lead to the sum rule~(\ref{eq:sumrule}).
For $p=1$ this relation exactly coincides with (\ref{E2cond}),
and implies the cancellation of the $E^2$ terms.

\subsection{Sum rule with boundary terms}

In the previous section, we have derived, neglecting some possible boundary terms, a sum rule
which ensures the cancellation of the terms that grow with the energy in the scattering amplitude.
We now would like to keep track of those boundary terms and see under which circumstances
the terms in the scattering amplitude that grow with energy still vanish.
As discussed above, the cancellation of the $E^4$ terms is always ensured by
the completeness of the eigenfunctions of $\partial_y^2$. However, the
sum rule in (\ref{E2cond}) will be modified if there are non-vanishing
boundary terms picked up when integrating by parts. The resulting corrections
to the sum rule relevant for the $E^2$ term is given by (we denote again by $[F]$ the boundary quantity
$F(\pi R) - F(0)$)
\begin{eqnarray}
\sum_{k} \, M_k^2\ \left( \int dy f_n^2(y) f_k^{\vphantom{2}}(y) \right)^2
= \sfrac{4}{3} M_n^2 \int dy f_n^4(y)
-\sfrac{2}{3} [f_n^3 f'_n]
\hphantom{spacespace}
\nonumber\\
        \label{E2bound}
-\sum_k \, [f_n^2 f'_k] \int dy f_n^2(y) f_k^{\vphantom{2}}(y)
+ 2  \sum_k \, [f_n^{\vphantom{2}} f'_n f_k^{\vphantom{2}}] \int dy f_n^2(y)
f_k^{\vphantom{2}}(y).
\end{eqnarray}
Thus one can see that (as expected) for arbitrary BC's
the $E^2$ terms no longer cancel. However, if one has pure
Dirichlet or Neumann BC's for all modes (some could have
Dirichlet and some could have Neumann, but none should have a mixed boundary
condition as in Eq. (\ref{mixedbc})) then all the extra boundary terms will vanish, and
thus the cancellation of the $E^2$ terms would remain. Clearly, the
terms in (\ref{E2bound}) involving $[f_n^{\vphantom{2}} f'_n]$  vanish, since for the
external mode either the function or its derivative should vanish on the
boundary. The one term that needs to be analyzed more carefully
is $\sum_k [f_n^2 f'_k] \int dy f_n^2(y) f_k(y)^{\vphantom{2}}$. If $f_n^{\vphantom{2}}$ satisfies
Neumann BC's, while $f_k$ Dirichlet, then the boundary piece
itself is not vanishing. However, in this case the expansion in terms of
$f_k$ converges even on the boundary, and thus $\sum_k f'_k(x) \int f_n^2(y)
f_k^{\vphantom{2}}(y) dy$ converges to $2 f_n^{\vphantom{2}}(x) f'_n(x)$, and thus we will
again have
a product $[f_n^{\vphantom{2}} f'_n]$ on the boundaries, which will vanish.

Thus we can conclude that in theories with only Dirichlet or Neumann
BC's imposed the $E^2$ terms in the scattering amplitudes will
always cancel. This implies, for example
that all models that are obtained via an
orbifolding procedure (that is taking an extra dimensional theory on a circle,
and projecting onto modes that have a given property under $y\to -y$ and
$\pi R -y \to \pi R +y$ projections) will be tree-unitary at least up to
around the
cutoff scale of the theory.   In fact, we can see that for all consistent
BC's from (\ref{bc}) the $E^2$ terms vanish. This shows that, as expected,
those BC's which follow from a gauge invariant Lagrangian will have a proper
high energy behavior. However, in the presence of mixed BC, the boundary terms
in the piece of the scattering amplitude that grows with $E^2$ are non-vanishing
and prevent the theory to be tree-unitarity up to the naive cutoff scale of the theory,
{\it e.g.}, the perturbative cutoff. The reason is that this corresponds to
adding an explicit mass term for the gauge bosons on the boundary which
violates gauge invariance. If however this comes from a gauge invariant
scalar kinetic term via the Higgs mechanism, then extra scalar
degrees of freedom need to be added.
As we will see explicitly in the example of
section~\ref{sec:BraneHiggs}, some extra degrees of freedom localized at the boundary
are indeed needed to restore tree-unitarity in this case.
In some limits, however, these extra localized states
can decouple without spoiling tree-unitarity.

We close this section by presenting the corrections to the sum rule
that is relevant for the evaluation of the $E^0$ terms of the scattering
amplitude:
\begin{eqnarray}
\sum_{k} \left( \int dy f_n^2 f_k^{\vphantom{2}} \right)^2 M_k^4
= \sfrac{16}{3} M_n^4 \int dy f_n^4
+\sfrac{28}{3} M_n^2 [f_n^3 f'_n]
-4 [f_n^{\vphantom{2}} {f'_n}^3]
+ \sum_k
[f_n^2 f'_k - 2 f_n^{\vphantom{2}} f'_n f_k^{\vphantom{2}}]^2
\nonumber\\
\vspace{-.5cm}
-\sum_k
\left( 4 M_n^2 [f_n^2 f'_k] \int dy f_n^2 f_k^{\vphantom{2}}
- 2   [{f'_n}^2  f'_k] \int dy f_n^2 f_k^{\vphantom{2}}
- 2 [f_n^2 f'_k] \int dy {f'_n}^2  f_k^{\vphantom{2}}
\right).\ \
\end{eqnarray}
This is the relation that needs to be used if one were to try to find
the actual unitarity bounds on the finite pieces of the elastic scattering
amplitudes, which is beyond the scope of this paper.

\section{A 4D Effective Theory Without a Higgs}
\setcounter{equation}{0}
\setcounter{footnote}{0}

We have seen above that the necessary conditions for cancelling
the growing terms in the elastic scattering amplitudes are
\begin{eqnarray}
        \label{E4again}
g^2_{nnnn} &=& \sum_k  g_{nnk}^2, \\
        \label{E2again}
4   g_{nnnn}^2 M_{n}^2 &=& 3 \sum_k  g_{nnk}^2 M_{k}^2\, .
\end{eqnarray}
One can ask the question, in what kind of low-energy effective theories
could these conditions be possibly satisfied. Assuming, that there are
$N$ gauge bosons, one can clearly always satisfy the relations
(\ref{E4again})-(\ref{E2again}) for the first $N-1$ particles,
as long as $g_{nnnn}^2 \geq g_{nnn}^2$. Given the
couplings $g_{111}$ and $g_{1111}$ and the mass of the lightest
gauge boson $M_1^2$, the necessary couplings and masses for the second gauge
boson can be calculated from (\ref{E4again})-(\ref{E2again}) to be
\begin{equation}
g_{112}^2=g_{1111}^2-g_{111}^2, \ \ M_2^2= \frac{M_1^2}{g_{112}^2}
\left( \frac{4}{3} g_{1111}^2-g_{111}^2 \right),
\end{equation}
assuming that a single extra gauge boson is needed to cancel the growing
amplitudes. Clearly, there are many other possibilities as well to satisfy
these equations for the $N-1$ lightest particles. However, the equation can
not be satisfied for the heaviest mode itself. The reason is that for the
heaviest particle one would get the equations
\begin{equation}
g_{NNNN}^2 =\sum_{k} g_{NNk}^2, \ \ \ g_{NNNN}^2= \sum_k g_{NNk}^2
\, \frac{3 M_k^2}{4 M_N^2}.
\end{equation}
Since the ratio $3 M_k^2/(4 M_N^2)<1$ for the heaviest mode, these
equations can not be solved no matter what the solution for the first $N-1$
particles was, as long as one has a finite number of modes. Thus we can see
that there can not be a heaviest mode, for every gauge boson there needs
to be one that is more heavy if one wants to ensure unitarity for all
amplitudes. Thus a Kaluza-Klein type tower must be present by these
considerations if the theory is to be fully unitary without any scalars, and
(\ref{E4again})-(\ref{E2again}) can be satisfied only in the presence of
infinitely many gauge bosons.

{}From an effective theory point of view one should however consider the
theory with a cutoff scale. Then there would be a finite number of modes
below this cutoff scale, for all of which one could ensure the unitarity
relations, except for the mode closest to the cutoff, for which one has
to assume that the UV completion plays a role in unitarizing the amplitude.
However, from this point of view we can see that the cutoff scale could be
much higher than the one naively estimated from the growing $E^2$ amplitudes
within the standard model. For example, in the minimal case a single
$W'$ and $Z'$ gauge bosons are necessary to unitarize the
$W_L^+W_L^-$ and $Z_LZ_L$ scattering amplitudes. In fact, the relations
(\ref{E4again})-(\ref{E2again}) can be satisfied with a $W'$ and $Z'$
that are heavy enough and sufficiently weakly coupled, so that their effects
would not have been observed at the Tevatron nor would they have significantly
contributed to electroweak precision observables.

For example, in the SM the scattering $W_L^+W_L^-\to W_L^+W_L^-$ is mediated
by $s$, $t$ and $u$-channel $Z$ and $\gamma$ exchange, and by the direct quartic
coupling. The cancellation of the $E^4$ term in the SM is ensured by the
relations
\begin{equation}
g^2_{WWWW}= g^2, \ \ \ g^2_{WWZ}= g^2 \cos^2 \theta_W, \ \ \
g^2_{WW\gamma} = e^2 = g^2 \sin^2 \theta_W.
\end{equation}
Let us now assume that the relation between $g^2_{WWWW}$ and $g^2_{WWZ}$ is
slightly modified due to the existence of a heavy $Z'$, which has a small
cubic coupling with the $W$'s $g^2_{WWZ'}$. The values of the three gauge
boson couplings $g_{WWZ}$ and $g_{WW\gamma}$ are not strongly constrained
by experiments, they are known to coincide with the SM values
to a precision of about 3-5\% , while there is basically no existing
experimental constraint on $g^2_{WWWW}$. Let us therefore assume that
the three gauge boson coupling $g^2_{WWZ}$ is smaller by a percent than
the SM prediction. Then in order to maintain the cancellation of the
$E^4$ terms one needs $g^2_{WWZ'} \sim 0.01 g^2_{WWZ}$. The cancellation of the
$E^2$ terms would then fix the mass of the $Z'$ to be about $560$ GeV.

Note that one would not expect a quartic $Z$ coupling or a $ZZ\gamma$
coupling, since there is none in the SM, thus there are no contributions
to $Z_LZ_L\to Z_LZ_L$ (contrary to the SM where the Higgs exchange provides
a finite term). However, to unitarize the $W_LZ_L\to W_LZ_L$ amplitude there
also needs to be a heavy $W'$, with couplings to $W$ and $Z$ $g^2_{WZW'}$
similar to $g^2_{WWZ'}$. Thus (without actually having calculated the
coefficients of the amplitudes for the inelastic scatterings) we expect that
$W'$ would also have a mass of order 500 GeV. This way all the amplitudes
involving only SM $W$'s and $Z$'s in initial and final states would be
unitarized. However, due to the sum rule explained above the scattering amplitudes
involving the $W'$ and $Z'$ in initial and final states can not all
be cancelled by only these particles. We expect that these
amplitudes would become large at scales that are set by the masses of the
$W'$ and $Z'$ rather than the ordinary $W$ and $Z$, so this would
roughly correspond to a scale of order $4\pi M_{W'}/g \sim 10$ TeV, were the
theory with only these particles would break down.

Since the 4D effective theory below $\Lambda=10$ TeV does not have an $SU(2)_L$ gauge
invariance (not even a hidden gauge invariance) there is an issue with the mass
renormalization for the gauge bosons: there will be quadratic divergences in the
$W$ and $Z$ mass.  (If the theory is completed into an extra dimensional
gauge theory with gauge symmetry breaking by boundary conditions,
then the masses will not really be quadratcially divergent, the divergences
will be cut off roughly by the radius of the extra dimension.)
However the theory is still technically natural since there is
a limit where (hidden) gauge invariance is restored and the divergence must
be proportional to the gauge invariance breaking spurion.  In the present example
we can estimate the divergent contribution to the $Z$ mass to be at most
\beq
\delta M_Z^2
\approx 0.01 \frac{ g^2_{WWZ}}{16 \pi^2} \Lambda^2
\approx 0.7\,  {\rm GeV}^2~.
\eeq

The three gauge boson couplings required between the heavy and the light gauge
bosons is quite small. Therefore loop contributions to the oblique parameters
$S,T,U$ will be strongly suppressed. However, one has to still ensure that
no large tree-level effects appear due to mixings between the heavy and
light gauge bosons, which could spoil electroweak precision observables.
In practice, this will probably require that the interactions of the
heavy $W'$ and $Z'$ obey a  global  SU(2)$_C$ custodial
symmetry which ensures $\rho =1$ in the SM

The masses of order 500 GeV would fall
to the borderline region of direct observation, assuming that their
couplings to fermions is equal to the SM couplings. However, if the
coupling to fermions is just slightly suppressed, or it decays dominantly to quarks or
through cascades,
 then the direct search
bounds could be easily evaded.

One should ask what kind of theories would actually yield low-energy
effective theories of the sort discussed in this section. We will see below
that some extra dimensional models could come quite close, though the masses
of the gauge bosons are lower than 500 GeV, while the $\rho$ parameter
has to be tuned to its experimental value. Before moving on to these models,
let us discuss whether any 4D theories with product group structure could
result in cancellations of the growing amplitudes. These models would
correspond to deconstructed extra dimensional theories. It has been shown
in \cite{Sekhar2} that for finite number $N$ of product gauge groups
there will be leftover pieces in the $E^2$ amplitude (without including Higgs
exchange) that scale with $N$ like $1/N^3$ (see Appendix~\ref{app:dec} for details).
Therefore, in any linear or
non-linear realizations of product groups broken by mass terms on the link
fields the amplitudes will not be unitary. However, already for $N=3$
one would get a suppression factor of 27 in the
$E^2$ piece of the scattering
amplitude, which would mean that the scale at which unitarity violations
would become visible would be of order 8 TeV, rather than the usual
1.5~TeV scale. Below we will focus on the models based on extra dimensions,
where the cancellation of the $E^2$ terms is automatic without a scalar
exchange.

\section{Examples of Unitarity in Higher Dimensional Theories}
\setcounter{equation}{0}
\setcounter{footnote}{0}

In this section we will present two examples of how the unitarity relations
(\ref{E4cancellation})-(\ref{E2cancellation}) are satisfied in particular extra dimensional theories.
The first example will be based on pure Dirichlet or Neumann boundary
conditions, where the cancellation as expected is automatic,
while the second example will involve mixed BC's for some of
the gauge bosons, which will imply the non-cancellation of the $E^2$
terms. However, as we will show unitarity can be restored by introducing
a boundary Higgs field.

\subsection{SU(2)$\to$ U(1) by BC}
\label{sec:orbifold}

Let us consider a 5D $SU(2)$ Yang--Mills theory compactified on an orbifold that leaves only a $U(1)$ subgroup
unbroken. We will further impose a Scherk--Schwarz periodic condition in order to project out all the
4D scalar fields coming from the component of the gauge fields along the extra dimension. The 5D gauge fields
thus satisfy:
\begin{eqnarray}
A_\mu (x,-y) = P A_\mu (x,y) P^{-1}
\ \ \ &{\rm and }& \ \ \
A_5 (x,-y) = - P A_5 (x,y) P^{-1},
\label{SS}
 \\
 A_\mu (x,y+2\pi R) = T A_\mu (x,y) T^{-1}
\ \ \ &{\rm and }& \ \ \
 A_5 (x,y+2\pi R) = T A_5 (x,y) T^{-1}.
\end{eqnarray}
where $A_M=A_M^a \tau^a/2$ ($\tau^a, a=1\ldots3$ are the Pauli matrices) is the gauge field and
$P=\rm{diag} (1,-1)$, the orbifold projection and $T=P$, the Scherk--Schwarz shift.
Note that the orbifold projection
and the  Scherk--Schwarz shift satisfy the consistency relation~\cite{SSorbifold}: $PTP=T^{-1}$.
This setup could alternatively also be described by two
$Z_2$ projections $y\to -y$ and $\pi R -y \to \pi R+y$. The action of the first
$Z$ is just as in (\ref{SS}), while the second one is similar with
$P$ replaced by $PT=P^2=1$.

Equivalently, we can describe this orbifold by a finite interval ($y \in [0,\pi R]$) supplemented by
the BC's:
\begin{eqnarray}
A^{1,2}_\mu (x,0)  = 0,
\partial_5A^3_{\mu} (x,0) = 0,
\partial_5 A^{a}_\mu (x,\pi R) = 0
 \\
\partial_5 A^{1,2}_5 (x,0) = 0,
 A^3_5 (x,0) = 0,
 A^a_5 (x,\pi R) = 0
\end{eqnarray}
Note that there is no zero mode coming from the fifth components of the gauge fields.
Therefore, we go to the unitary gauge which is the same as the axial gauge ($A_5=0$).
Then the KK decomposition is
\begin{eqnarray}
        \label{eq:SinKA1}
& \displaystyle
A^{1}_\mu (x,y)  =\sum_{k=0}^{\infty} \frac{1}{\sqrt{\pi R}}
\sin \left(   \frac{(2k+1) y}{2R} \right) (W^{+\, (k)}_\mu (x) + W^{-\, (k)}_\mu (x) ) ,
\\
        \label{eq:SinKA2}
& \displaystyle
A^{2}_\mu (x,y)  =\sum_{k=0}^{\infty} \frac{i}{\sqrt{\pi R}}
\sin \left(   \frac{(2k+1) y}{2R} \right) (W^{+\, (k)}_\mu (x) - W^{-\, (k)}_\mu (x) ) ,
\\
        \label{eq:CosK}
&  \displaystyle
A^{3}_\mu (x,y) = \sum_{k=0}^{\infty} \sqrt{\frac{2}{2^{\delta_{k,0}}\pi R}}
\cos \left(   \frac{k y}{R} \right) \gamma^{(k)}_\mu (x).
\end{eqnarray}
The spectrum contains a massless photon, $\gamma^{(0)}$, and its KK excitations,
$\gamma^{(k)}$, of mass  $M_{\gamma^{(k)}}=k/R$   as well as a
tower of massive charged gauge bosons, $W^{\pm\, (k)}$,
of mass $M_{W^{(k)}}=(2k+1)/(2R)$. With the above wavefunctions, it is easy to
explicitly compute the cubic and quartic effective couplings and check
the general sum rules of Section~\ref{sec:general}. For instance, for
the elastic scattering of  $W$'s, the relevant couplings are:
\begin{eqnarray}
& \displaystyle
g_{W^{(n)}W^{(n)}\gamma^{(k)}} = \frac{g_5}{2\sqrt{\pi R}}
\left( \delta_{k,0} - \frac{1}{\sqrt{2}} \delta_{k,2n+1}\right),
\\
& \displaystyle
g^2_{W^{(n)}W^{(n)}W^{(n)}W^{(n)}} = \frac{3 g_5^2}{8 \pi R}.
\end{eqnarray}
The BC's conserve KK momenta up to a
sign and therefore only $\gamma^{(0)}$ and $\gamma^{(2n+1)}$
can contribute to the elastic scattering of $W^{(n)}$'s.
The sum rules~(\ref{E4cancellation})--(\ref{E2cancellation}) are trivially fulfilled.

The point of this section was to show that it is indeed possible
to give a mass to gauge boson without relying on a Higgs mechanism
to restore unitarity. The orbifold symmetry breaking mechanism illustrated with this example is
however restrictive since it uses a $Z_2$ symmetry of the action and in the simplest cases
(abelian orbifolds using an inner automorphism) it is even impossible to reduce
the rank the gauge group, which is a serious obstacle to the construction
of realistic phenomenological models.
There are, however, more general BC's we can impose
that are not equivalent to a simple
orbifold compactification but still lead to a well behaved effective theory.

\subsection{Completely broken SU(2) by mixed BC's
and the need for a boundary Higgs field}
        \label{sec:BraneHiggs}

In the previous section, we considered a breaking of $SU(2)$ down to $U(1)$ by an orbifold compactification.
We have shown that, in agreement with our general proof of Section~\ref{sec:general}, the scattering amplitude
of the gauge bosons that acquire a mass through compactification has a good high energy behavior.
The cancellation of the terms growing with  energy is ensured by the exchange of higher massive gauge bosons
and does not require the presence of any scalar field. We would like now to consider some more general boundary
conditions than the ones coming from orbifold compactification. In this case, we will see that the boundary terms
of the sum rule~(\ref{E2bound}) are non vanishing and thus lead to a violation of the unitarity at tree level, however
unitarity can be restored by a scalar field living at the boundary. We illustrate these results with an example
of an $SU(2)$ completely broken by mixed (neither Dirichlet nor Neumann) BC's.\footnote{Ref.~\cite{NW}
presented a model of GUT breaking using mixed BC's.}

Let us thus now consider the BC's:
\begin{eqnarray}
        \label{eq:mixedBC}
&
\partial_5 A^{a}_\mu (x,0)  = 0, \ \ \
\partial_5 A^{a}_\mu (x,\pi R) = V   A^{a}_\mu (x,\pi R),
 \\
&
A^{a}_5 (x,0) = 0, \ \ \
A^{a}_5 (x,\pi R) = 0.
\end{eqnarray}
A general solution can be decomposed on the KK basis
\begin{equation}
A^{a}_\mu (x,y)  =\sum_{k=1}^{\infty} f_k(y)   A^{(k)}_\mu (x),
\ \ {\rm with } \ f_k(y)=\frac{a_k}{\sin (M_k \pi R)} \cos (M_k y).
\end{equation}
The BC at the origin, $y=0$, is trivially satisfied while the condition at $y=\pi R$
determines the mass spectrum through the transcendental equation:
\begin{equation}
M_k \tan (M_k \pi R) = - V.
\end{equation}
The parameter $V$ controls the gauge symmetry breaking: when $V=0$, the BC's
are those of a orbifold compactification with no symmetry breaking and when $V$ is turned on there is no
zero mode any more and the full $SU(2)$ gauge group is broken completely. Note that there is no zero mode
either for the fifth component of the gauge field and thus $A_5$ can be gauged away, leaving no scalar field
in the low energy 4D effective theory.

The normalization factor $a_k$ is determined by requiring that the KK modes are canonically normalized,
$\int_0^{\pi R}f_k^2(y)=1$, leading to
\begin{equation}
a_k = \frac{\sqrt{2}}{\sqrt{\pi R (1+  M_k^2/V^2)-1/V}}.
\end{equation}
Note that this KK decomposition can equivalently be obtained through
a much more lengthy procedure consisting in two steps (see~\cite{Muck} for details):
({\it i}) assume $V=0$, get a KK decomposition as in~(\ref{eq:CosK}) and obtain the
corresponding effective action;
({\it ii})  reintroduce the parameter $V$ in the effective action as a mass term that mixes
all the previous modes and the true eigenmodes are obtained by diagonalizing the corresponding infinite
mass matrix.

Unlike in the case of the previous section involving only Dirichlet or Neumann BC's,
the mixed BC's~(\ref{eq:mixedBC}) give rise to boundary terms in the sum rule~(\ref{E2bound})
needed for the computation of the scattering amplitude. We obtain:
\begin{equation}
\sum_{k} \, M_k^2\ \left( \int dy f_n^2 f_k^{\vphantom{2}} \right)^2
= \sfrac{4}{3} M_n^2 \int dy f_n^4
+\sfrac{1}{3} V f_n^4(\pi R).
\end{equation}
Consequently, the scattering amplitude has a residual piece growing with the energy square
which is proportional to the order parameter, $V$:
\begin{equation}
A^{(2)}=
\frac{i g_5^2}{M_n^2} V f_n^4(\pi R)
\left( - \delta^{ab}\delta^{cd} +    \delta^{ac}\delta^{bd}\,  \sin^2 \theta/2
+    \delta^{ad}\delta^{bc} \, \cos^2 \theta/2 \right).
\end{equation}
Besides the case $V=0$, there is another limit where these boundary terms actually
vanish. Indeed, when $V\gg 1/R$, the low-lying eigenmasses can be approximated by
\begin{equation}
M_k \sim \frac{2k+1}{2R} \left(1+\frac{1}{\pi R V}+\ldots \right),
\ \ k=0,1,2\ldots
\end{equation}
and the normalization factor is $a_k \sim 1/\sqrt{\pi R}$, thus
\begin{equation}
 V f_n^4(\pi R) \sim \frac{(2n+1)^4}{4 \pi^2 R^{6} V^3}.
\end{equation}
Therefore, the terms in the scattering amplitude that grow with the energy do cancel
when $V$ goes to infinity. The physical reason of such a cancellation is clear: in the large
$V$ limit, the brane localized mass term becomes large and the wavefunctions are expelled from the brane
thus the mixed BC $\partial_5 f_k (\pi R)=V f_k (\pi R)$ in the limit
$V\to \infty$ simply becomes equivalent to the Dirichlet BC $f_k (\pi R)=0$ which,
as we already know, does not lead to any unitarity violation.

\begin{figure}[ht]
\centerline{\includegraphics[width=0.75\hsize]{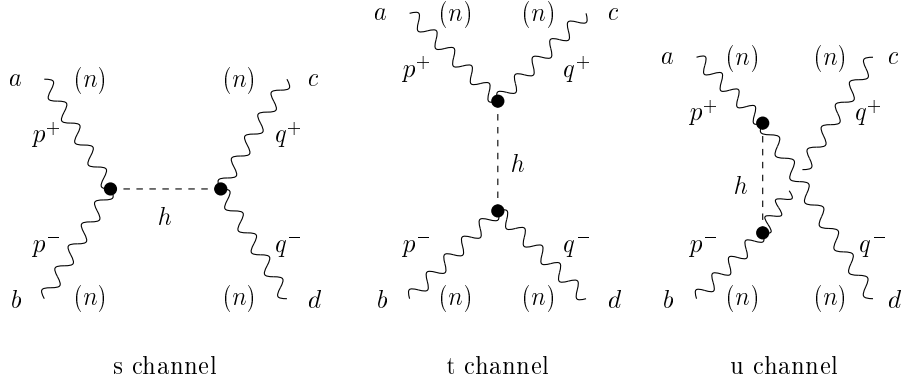}}
\label{fig:Higgs}
\caption{Higgs exchange contributing at tree level
to the gauge boson elastic scattering amplitude. The Higgs boson is localized at $y=\pi R$ and
the vertex Higgs-Gauge-Gauge is proportional to $f_n^2(\pi R)$.}
\end{figure}

As we already said, a finite and non-vanishing order parameter $V$ can be interpreted as
an explicit gauge boson mass term localized at the boundary.  This explicit breaking can be UV completed
using the usual Higgs mechanism. In our $SU(2)$ model, we can introduce a 4D $SU(2)$ doublet
localized at $y=\pi R$. Giving a vacuum expectation value to the lower component of the localized Higgs doublet
\begin{equation}
\langle H \rangle =
\frac{1}{\sqrt{2}}
\left(
\begin{array}{c}
0 \\
v
\end{array}
\right)
\end{equation}
gives rise to the mass term $\int d^4x \sfrac{1}{8}g_5^2 v^2 A_\mu^2 (x,\pi R) $ in the 5D action.
To obtain the BC (see Appendix~\ref{app:braneHiggs} for technical details) we need to
vary the action that contains both a bulk and a boundary piece.
The relevant part of the action is
\begin{equation}
\left( \int_0^{\pi R}dy\,   \sfrac{1}{2} \partial_5 A_\mu \partial_5 A^\mu \right)
+ \sfrac{1}{8}g_5^2 v^2 A^2_{\mu\, |\pi R}.
\end{equation}
Varying the bulk terms will introduce, after integration by parts, the usual equation of motion as well
as  boundary terms:
\begin{equation}
-\left( \int_0^{\pi R} dy\,   \partial_5^2 A_\mu \,  \delta A^\mu \right)
+\left( \partial_5 A_\mu \, \delta A^\mu +  \sfrac{1}{4}g_5^2 v^2 A_\mu   \, \delta A^\mu \right)_{|\pi R}~.
\end{equation}
The boundary terms thus impose the mixed BC:
\begin{equation}
        \label{eq:BCfromHiggs}
\partial_5 A_\mu^a (x,\pi R) = V A_\mu^a (x,\pi R),
\ \ {\rm with }\ \
V= - \sfrac{1}{4} g_5^2 v^2.
\end{equation}
The three Goldstone bosons are eaten by the KK gauge bosons that would be massless when $V=0$
and we are left with only one physical real scalar field, the Higgs boson. At tree level, the exchange
of this Higgs boson also contributes to the gauge boson scattering amplitude through the diagrams depicted in
Fig.~\ref{fig:Higgs}. The Higgs being localized at  $y=\pi R$, its coupling to
the gauge bosons is proportional to $f_k^2 (\pi R)$. Thus we get a contribution to the scattering amplitude
that grows with the energy square:
\begin{equation}
A^{(2)}=
\frac{i g_5^4 v^2}{4 M_n^2} f_n^4(\pi R)
\left( - \delta^{ab}\delta^{cd} +    \delta^{ac}\delta^{bd}\,  \sin^2 \theta/2
+    \delta^{ad}\delta^{bc} \, cos^2 \theta/2 \right).
\end{equation}
Using the relation~(\ref{eq:BCfromHiggs}), we get that (as expected)
the Higgs exchange exactly cancels
the terms in scattering amplitude from the gauge exchange  that grows like $E^2$.
Remarkably enough, the Higgs boson can be decoupled ($v \to \infty$ limit) without
spoiling the high energy behavior of the massive gauge boson scattering.
Note, that this limit would simply correspond to the $A_\mu^a=0$ BC from
(\ref{bc}).

\section{A Toy Model for Electroweak Symmetry Breaking via BC's}
        \label{sec:LR}
\setcounter{equation}{0}
\setcounter{footnote}{0}

We want to study in this section the possibility to break the electroweak symmetry $SU(2)_L\times U(1)_Y$
down to  $U(1)_Q$ along the lines of the previous section, {\it i.e.}, by BC's without
relying on a Higgs mechanism either in the bulk or on the brane (a realistic model
of electroweak symmetry breaking without  any fundamental scalar field has been constructed in~\cite{us}, see
also~\cite{A5Higgs}.
Here we want to go further and totally remove any 4D scalar fields).
The first idea would be to extend the analysis
of Section~\ref{sec:BraneHiggs} to include a mixing between  $SU(2)$ and $U(1)$. However, considering
the limit $V\to \infty$, we would get a mass degeneracy for the $W^\pm$ and the $Z$ gauge bosons.
Our point is to show that we can actually lift this degeneracy and obtain a spectrum that depends
on the gauge couplings. We do not claim to have a fully realistic model but we want to construct a  toy model
that has the characteristics of the Standard Model without the Higgs and that remains theoretically consistent.

Let us consider $SO(4)\times U(1)_{\BL}\sim SU(2)_L\times SU(2)_R \times U(1)_{\BL}$
compactified on an interval $[0,\pi R]$ (for other models using
a left-right symmetric extra dimensional bulk see~\cite{otherleftright},
thought the breaking patter of symmetries is very different in those models).
At one end, we break $SO(4)$ down to $SU(2)_D$  by Neumann and Dirichlet BC's.
At the other end of the interval, we break $SU(2)_R\times U(1)_{\BL}$ down to $U(1)_Y$
by mixed BC's and we will consider the limit $V\to \infty$ in order to ensure
unitarity without having to introduce extra scalar degrees of freedom (alternatively, one can
directly impose the equivalent Dirichlet BC's, however, the gauge structure is more
transparent using the limit of mixed BC's). Thus only $U(1)_Q$ remains unbroken.
We denote by
$A^{R\,a}_{M}$, $A^{L\,a}_{M}$ and $B_M$ the gauge bosons of
$SU(2)_R$, $SU(2)_L$ and $U(1)_{\BL}$ respectively; $g$ is the gauge coupling
of the two $SU(2)$'s and ${g'}$, the gauge coupling of the $U(1)$. We  consider the two linear combinations\footnote{
It can be seen that every other linear combinations, except the identity, would have not maintain
any gauge invariance. The particular combination chosen here preserves, at the boundary,
an $SU(2)_D$ gauge invariance.}
of the $SU(2)$ gauge bosons
$A^{\pm\,a}_{M}=\sfrac{1}{\sqrt{2}} (A^{L\,a}_M \pm A^{R\,a}_M)$.
We impose the following BC's:
\begin{eqnarray}
&
{\rm at }\  y=0:
&
\left\{
\begin{array}{l}
\partial_5 A^{+\,a}_\mu = 0, \
A^{-\,a}_\mu  = 0, \
\partial_5 B_\mu = 0,
\\
\tv{15}
A^{+\,a}_5 = 0, \
\partial_5 A^{-\,a}_5  = 0, \
B_5 = 0.
\end{array}
\right.
\label{bc1}\\
&
{\rm at }\ y=\pi R:
&
\left\{
\begin{array}{l}
\partial_5 A^{L\,a}_\mu=0, \
\partial_5 A^{R\,1,2}_\mu = -\sfrac{1}{4}  g^2  v^2 A^{R\,1,2}_\mu,
\\
\tv{15}
\partial_5 B_\mu = -\sfrac{1}{4}  {g'} v^2({g'} B_\mu - g A^{R\,3}_\mu), \
\partial_5 A^{R\,3}_\mu = \sfrac{1}{4}  g v^2({g'} B_\mu - g A^{R\,3}_\mu),
\\
\tv{15}
A^{L\,a}_5=0, \ A^{R\,a}_5=0, \ B_5 = 0.
\end{array}
\right.
\label{bc2}
\end{eqnarray}
One has to check now, that the equations (\ref{bc}) are indeed satisfied
requiring these BC's. We will only explain how the first
of (\ref{bc}) is satisfied at $y=0$, one can similarly check all the other
conditions. From $A^{-\,a}_\mu  = 0$ we get that $\delta A^{L a}_\mu =
\delta A^{R a}_\mu$. Thus we need to show that $F^{L a}_{\nu 5}+
F^{R a}_{\nu 5}=0$, which is true due to the requirements
$\partial_5 A^{+\,a}_\mu = 0$, $A_\mu^{-\,a}=0$, and $A^{+\,a}_5 = 0$. Note, that these
BC's are a non-trivial example of (\ref{bc}) being satisfied
without a term-by-term cancellation of the actual boundary variations,
but rather by a cancellation among the various terms.

One may wonder where these complicated looking BC's
originate from. In fact, they correspond to a physical situation where
one has an orbifold projection based on an outer automorphism
$SU(2)_L \leftrightarrow SU(2)_R$ around one of the fixed points.
Indeed the BC's can be seen as deriving from the orbifold projections
($\hat{y}=y+\pi R$):
\begin{eqnarray}
&
A^{L\, a}_\mu (x, -y) = A^{R \, a}_\mu (x, y) \, , \
A^{R\, a}_\mu (x, -y) = A^{L \, a}_\mu (x, y) \, , \
B_\mu (x, -y) = B_\mu (x, y)\ , \\
&
A^{L\, a}_\mu (x, -\hat{y}) = A^{L\, a}_\mu (x, \hat{y}) \, , \
A^{R \, a}_\mu (x, -\hat{y}) = A^{R\, a}_\mu (x, \hat{y}) \, , \
B_\mu (x, -\hat{y}) = B_\mu (x, \hat{y})  \   .
\end{eqnarray}
The projections on the fifth components of the gauge fields are the same except an additional
factor $-1$.
At the $y=\pi R$ end point,
a localized $SU(2)_R$ scalar doublet of $U(1)$ charge $1/2$ acquires a VEV  $(0,v)/\sqrt{2}$
and breaks $SU(2)_R \times U(1)$ down to $U(1)_Y$. The mass terms is
responsible for the mixed BC's~(\ref{bc2}).

For a finite VEV, the 4D Higgs scalar localized at $y=\pi R$ is needed to keep
the theory unitary; we will however consider the limit $v \to \infty$ where this scalar field
can be decoupled without spoiling the high energy behavior of the gauge boson scattering.

Due to the mixing of the various gauge groups, the KK decomposition is more involved
than in the simple example of Section~\ref{sec:BraneHiggs} but it is obtained
by simply enforcing the BC's (we denote by $A^{L,R\, \pm}_\mu$ the linear
combinations
$\sfrac{1}{\sqrt{2}} (A^{L,R\, 1}\mp i A^{L,R\, 2})$):
\begin{eqnarray}
        \label{eq:KKB}
B_\mu (x,y) & = & g\,  a_0 \gamma_\mu (x)
+  {g'} \sum_{k=1}^{\infty} b_k \cos (M^{Z}_k y) \, Z^{(k)}_\mu (x)\, ,
\\
        \label{eq:KKAL3}
A^{L\, 3}_\mu (x,y) & = &
{g'} \, a_0 \gamma_\mu (x)
-  g \sum_{k=1}^{\infty}   b_k  \frac{\cos (M^{Z}_k (y-\pi R))}{2 \cos (M^{Z}_k \pi R)} \, Z^{(k)}_\mu (x) \, ,
\\
        \label{eq:KKAR3}
A^{R\, 3}_\mu (x,y) & = &
{g'}  a_0 \gamma_\mu (x)
- g  \sum_{k=1}^{\infty}  b_k \frac{\cos (M^{Z}_k (y+\pi R))}{2 \cos (M^{Z}_k \pi R)} \, Z^{(k)}_\mu (x) \, ,
\\
        \label{eq:KKALpm}
A^{L\, \pm}_\mu (x,y) & = &
\sum_{k=1}^{\infty}   c_k \cos (M^{W}_k (y-\pi R))\, W^{(k)\, \pm}_\mu (x) \, ,
\\
        \label{eq:KKARpm}
A^{R\, \pm}_\mu (x,y) & = &
\sum_{k=1}^{\infty} c_k \cos (M^{W}_k (y+\pi R))\, W^{(k)\, \pm}_\mu (x)  \, .
\end{eqnarray}
The normalization factors, in the large VEV limit, are given by
\begin{eqnarray}
        \label{eq:norm}
a_0  = \frac{1}{\sqrt{\pi R}} \frac{1}{\sqrt{g^2+2 {g'}^2}}\, , \
b_k   \sim   \frac{\sqrt{2}}{\sqrt{ \pi R}} \frac{1}{\sqrt{g^2+2{g'}^2}} \, , \
c_k  \sim \frac{1}{\sqrt{ \pi R}}\, .
\end{eqnarray}

The spectrum is made up of a massless photon, the gauge boson associated with the unbroken
$U(1)_Q$  symmetry, and some towers of massive charged and neutral gauge bosons,
$W^{(k)}$ and $Z^{(k)}$ respectively.
The masses of the $W^\pm$'s are solution of
\begin{equation}
M^W_k \tan (2 M^W_k \pi R) = \sfrac{1}{4} g^2 v^2 \, ,
\end{equation}
and in the large $v$ limit, we get the approximate spectrum:
\begin{equation}
M^W_k = \frac{2k-1}{4R} \left(1- \frac{2}{\pi g^2 v^2 R}+\ldots\right), \ \ k=1,2\ldots
\end{equation}
The masses of the $Z$'s are solution of
\begin{equation}
M^Z_k \tan (M^Z_k \pi R) = \sfrac{1}{8}(g^2+2{{g'}}^2)v^2 - \sfrac{1}{8} g^2 v^2 \tan^2 (M^Z_k \pi R) \, ,
\end{equation}
and in the large $v$ limit, we get two towers of neutral gauge bosons:
\begin{eqnarray}
M^Z_k = \left(M_0 + \frac{k-1}{R}\right)
\left( 1 - \frac{2}{(g^2+{{g'}}^2)v^2 \pi R} +\ldots\right), \ \   k=1,2\ldots
\\
M^{Z'}_k = \left(-M_0 + \frac{k}{R}\right)
\left( 1 - \frac{2}{(g^2+{{g'}}^2)v^2 \pi R} +\ldots\right), \ \   k=1,2\ldots
\end{eqnarray}
where $M_0=\sfrac{1}{\pi R} \arctan \sqrt{1+2 {{g'}}^2/g^2}$. Note that
$1/(4R)<M_0<1/(2R)$ and thus the $Z'$'s are heavier than the $Z$'s
($M^{Z'}_k>M^Z_k$). We also get that the lightest $Z$ is heavier than
the lightest $W$ ($M^Z_1>M^W_1$), in agreement with the SM spectrum.

The BC's break KK momentum conservation and as a consequence
all the KK will interact to each other. For instance, the cubic effective couplings between
the $W$'s and the $Z$'s (and the $Z'$'s) are, in the large VEV limit,
\begin{equation}
g_{W^{(n)} W^{(n)}Z^{(k)}} =
-\frac{2 g^2}{\sqrt{ \pi^3 R^3 (g^2+g^{\prime\, 2})}}
\frac{M_n^{W\, 2}}{(4M^{W\, 2}_n-M^{Z\, 2}_k)M^Z_k} \, ,
\end{equation}
while the couplings between the $W$'s and the photon are
\begin{equation}
g_{W^{(n)} W^{(n)}\gamma} =
\frac{g{g'}}{\sqrt{g^2+2 {g'}^2}} \frac{1}{\sqrt{\pi R}}\, .
\end{equation}

Let us now discuss how to introduce matter fields. Locally at the $y=0$ boundary, a
$SU(2)_D\times U(1)_{\BL}$ subgroup remains unbroken. We can introduce some matter fields
localized on this boundary. Consider a $SU(2)$ scalar doublet with a $U(1)$ charge $\sqrt{2}q$.
Its interactions with the gauge boson KK modes  are generated through the localized covariant derivative
\begin{equation}
D_\mu \Phi = \partial_\mu \Phi - \frac{i}{2}
\left(
\begin{array}{cc}
2 \sqrt{2} {g'} q B_{\mu}+ g A^{+\, 3}_\mu & g(A^{+\, 1}_\mu - i A^{+\, 2}_\mu) \\
 g(A^{+\, 1}_\mu + i A^{+\, 2}_\mu) &   2 \sqrt{2} {g'} q B_{\mu}- g A^{+\, 3}_\mu
 \end{array}
 \right)_{|0} \Phi  .
\end{equation}
Using the KK decomposition (\ref{eq:KKB})-(\ref{eq:KKARpm}), we evaluate
the value of the gauge fields at the boundary and the scalar covariant derivative becomes
\begin{eqnarray}
\hspace*{-0.1cm}D_\mu \Phi =
\partial_\mu \Phi
&-& i\sqrt{2}g{g'} a_0
\left(
\begin{array}{cc}
q+\sfrac{1}{2} & 0\\
0 & q-\sfrac{1}{2}
\end{array}
\right)
\gamma_\mu \Phi
%
- \sum_{k=1}^{\infty}
\frac{i b_kg^2}{\sqrt{8}}
\left(
\begin{array}{cc}
\sfrac{4q {g'}^2}{g^2} - 1  & 0\\
0 &  \sfrac{4q{g'}^2}{g^2} + 1
\end{array}
\right)
Z^{(k)}_\mu  \Phi
\nonumber \\
&-& \sum_{k=1}^{\infty}
i g c_k \cos (M^W_k \pi R)
\left(
\begin{array}{cc}
0  & W^{(k)\, +}_\mu \\
W^{(k)\, -}_\mu &  0
\end{array}
\right)
\Phi.
\end{eqnarray}
The interactions between the scalar doublet $\Phi$ and the first massive KK gauge bosons $Z^{(1)}$ and $W^{(1)}$
will exactly reproduce the SM interactions between a $SU(2)_L$ doublet of hypercharge
$q$ and the $Z$ and the $W$'s provided that the normalization factors $a, b_1, c_1$ satisfy
\begin{equation}
b_1=\sqrt{2} a_0,\quad b_1 = \frac{2 c_1 \cos (M^W_1 \pi R)}{\sqrt{g^2+2{g'}^2}}.
\end{equation}
In the infinite VEV limit, from the expressions~(\ref{eq:norm}) it can be checked that these
relations are exactly satisfied and the 4D SM couplings are expressed in terms of the 5D gauge couplings by
\begin{equation}
g_{4D}=\frac{g}{\sqrt{\pi R}},
\ \
{g'}_{4D}=\frac{\sqrt{2} {g'}}{\sqrt{\pi R}}.
\end{equation}
In the same way the SM $SU(2)_L$ singlets will
correspond to $SU(2)_D$ singlets charged under $U(1)$ and localized at the $y=0$ boundary.
When the corrections to the normalization factors for a finite VEV are included,
the interactions between the matter and the gauge bosons do not reproduce exactly the structure of
the SM. It has also to be noted that in the infinite VEV limit the normalization factors
are independent of the KK level, which means that the couplings of matter to higher KK states
will be unsuppressed.

One can also try to identify the couplings of matter localized at the $y=\pi R$ end point.
In particular, it can be seen that the lowest component of a $SU(2)_R$ doublet of $U(1)$ charge
$1/2$ does couple, in the $v\to \infty$ limit, to none of the gauge bosons. This explains why the
localized Higgs boson does not contribute to restore unitarity in the massive gauge boson scattering.

This toy model resembles the SM in that the lowest lying KK modes
of the gauge bosons have masses similar to the $\gamma ,W$ and $Z$, and the
couplings of the brane localized fields can be made equal to the couplings
of the SM fermions. However, there are clearly several reasons why this
particular model is not realistic.

The first reason is the $M_W/M_Z$ mass ratio. Even though we do get masses that
depend on the gauge couplings, which is a quite non-trivial step forward,
nevertheless the ratio does not exactly agree with the SM prediction. In the
$v \to \infty$ limit the ratio becomes
\begin{equation}
\frac{M_W^2}{M_Z^2}=\frac{\pi^2}{16}\arctan^{-2}
\sqrt{1+\frac{g_{4D}^{' 2}}{g_{4D}^2}}
\sim 0.85,
\end{equation}
and hence the $\rho$ parameter is
\begin{equation}
\rho=\frac{M_W^2}{M_Z^2 \cos^2 \theta_W}\sim 1.10 \ .
\end{equation}
Thus the mass ratio is close to the SM value, however the ten percent deviation
is still huge compared to the experimental precision. It is possible to get a more realistic value
of the $\rho$ parameter by keeping a finite VEV $v$. The price to pay is that the coupling of matter
to the gauge boson will not exactly reproduce the structure of the SM couplings: for instance,
if we match the coupling to the photon and to the $W$'s, we will get a deviation
of order $M_W^2/v^2$ in the coupling to the $Z$. We can simply estimate the
value of $v$ needed in order to get $\rho=1$:
\begin{equation}
\frac{M_W^2}{M_Z^2}
\sim \frac{\pi^2}{16}\arctan^{-2}
\sqrt{1+\frac{g_{4D}^{' 2}}{g_{4D}^2}}
\left( 1 - 32 (\delta M_0 - \delta M_W + \delta M_Z) \frac{M_W^2}{v^2} \right)
\end{equation}
where $\delta M_0,  \delta M_W, \delta M_Z$ are complicated functions
of the 4D gauge couplings.
The mass ratio can be tuned to exactly coincide with the experimental value, as long as
$v$ is lowered to about $v\sim 640$~GeV.  However, we can
see that a realistic mass ratio would require quite a low value of $v$, which
would imply that the scalar localized at $y=\pi R$ has a significant
contribution to the scattering amplitude. Also, for finite values of
$v$ one can no longer match up all the couplings of the brane localized fields
to their SM values.

The next issue is the masses of the KK excitations of the $W$ and $Z$. From the
expression $M^W_k \sim (2k-1)/(4R)$ one can see that $M^W_2 \sim 3 M^W_1
\sim 240$ GeV. This is too low if the coupling to the SM fermions is not
suppressed (as would be the case for brane localized fermions discussed
above). The third issue is that with brane localized fermions of the sort that
we discussed above, it is not possible to give the fermions a mass. The SM
Higgs serves two purposes: to break electroweak gauge symmetry and to give masses
to the SM fermions. We have eliminated the Higgs and broken electroweak gauge
symmetry by BC's. In order to be able to write down
fermion masses one would have to include them into the theory in a different
manner.

In order to get a more realistic theory, we need to modify the structure
of the model. For fermion masses, putting the fermions into the bulk and
only couple them to $SU(2)_L$ should be sufficient. Other possible
modifications are to put the Higgs that breaks $SU(2)_R\times U(1)_{B-L}$
in the bulk, or to consider warped backgrounds. Work along these directions is
in progress~\cite{inprogress}.

\section{Conclusions}
\setcounter{equation}{0}
\setcounter{footnote}{0}

We have investigated the nature of gauge symmetry breaking by boundary
conditions. First we have derived the consistent set of boundary conditions
that could minimize the action of a gauge theory on an interval. These
BC's include the commonly applied orbifold conditions, but there is a much
wider set of possible conditions. For example, it is simple to reduce
the rank of gauge groups. To find out more about the theories where
gauge symmetry breaking happens via BC's, we have investigated the high-energy
behavior of elastic scattering amplitudes. We have found, that for all
generic consistent BC's derived before the contributions to the amplitude
that would grow with the energy as $E^4$ or $E^2$ will always vanish, thus
these theories seem to have a good high-energy behavior just as gauge theories
broken by the Higgs mechanism would. However, since these are higher
dimensional theories, tree unitarity will still break down due to the
non-renormalizable nature (growing number of KK modes) of these models.

We have speculated, that perhaps the breaking of gauge symmetries via BC's
could replace the usual Higgs mechanism of the SM. We have shown an effective
field theory approach and a higher dimensional toy model for electroweak
symmetry breaking via BC's. Clearly there is still a long way to go to find
a fully realistic implementation of this new way of electroweak symmetry
breaking without a Higgs boson.

\section*{Acknowledgments}
We thank Nima Arkani-Hamed, Michael Chanowitz, Sekhar Chivukula,
Emilian Dudas, Josh Erlich,
Mary K. Gaillard, Michael Graesser, Antonio Masiero, Matthias Neubert,
Yasunori Nomura
and Matthew Schwartz
for useful discussions and comments.
C.C. and C.G. thank the Harvard High Energy Group for providing a
stimulating environment during a visit while we were working on this project.
C.G. is also grateful to the Particle Theory Group at Cornell University
and to the Theoretical Physics Group at the Lawrence Berkeley
Laboratory and UC Berkeley for their kind hospitality when part of this
work has been completed.
The research of C.C.
is supported in part by the DOE OJI grant DE-FG02-01ER41206 and in part by
the NSF grant PHY-0139738.
C.G. and L.P. are supported in part by the RTN European Program HPRN-CT-2000-00148
and the ACI Jeunes Chercheurs 2068. H.M. is supported in part by the United States Department of Energy,
Division of High Energy Physics under contract DE-AC03-76SF00098 and
in part by the National Science Foundation grant PHY-0098840.
J.T. is supported by the US Department of Energy under contract W-7405-ENG-36.


\appendix

\section*{Appendix}


\section{BC's for Gauge Theories with Scalars}
\label{app:bc}
\setcounter{equation}{0}
\setcounter{footnote}{0}

In this Appendix we continue the discussion of BC's
for gauge theories on an interval. First we will consider the case of a
gauge theory with bulk scalar fields, then a gauge theory with scalars
localized at the endpoints.

\subsection{Gauge theory with a bulk scalar}

Let us now discuss how the BC's and the bulk equations of
motion are modified in the presence of a bulk scalar field that
gets an expectation value. We will use the notation of \cite{PS} Chapter 21,
where all complex scalars are rewritten in terms of real components
denoted by $\Phi_i$, and expanded around their VEV's as
$\Phi_i=\langle \Phi_i \rangle +\chi_i$. The covariant derivative is
$D_M \Phi_i = \partial_M \Phi_i +g A_M^a T^a_{ij} \Phi_j$, where the
$T^a_{ij}$ generators are real and antisymmetric. The quadratic part of the
action is then given by
\begin{eqnarray}
        \label{eq:lagrbulkhiggs1}
&&\mathcal{S}=
\int d^4x\,\int_0^{\pi R} dy \left( -\frac{1}{4} F_{\mu\nu}^a F^{a\mu\nu}
-\frac{1}{2} F_{\mu 5}^a F^{a\mu 5}-
\frac{1}{2\xi} \left( \partial_\mu A^{a\mu}
-\xi (\partial_5 A_5^a+g F^a_i \chi_i)\right)^2 \right. \nonumber \\
&&\left. +\frac{1}{2} D_\mu \Phi_i D^\mu \Phi_i+
\frac{1}{2} D_5 \Phi_i D^5 \Phi_i -V(\Phi )\right).
\end{eqnarray}
Here we have added the modified form of the gauge fixing term. Expanding
this Lagrangian to quadratic order we get
\begin{eqnarray}
        \label{eq:lagrbulkhiggs}
&&\mathcal{S}=
\int d^4x\,\int_0^{\pi R} dy \left(
-\frac{1}{2} A_\nu^a (-\partial_\rho \partial^\rho g^{\mu\nu} +\partial^\mu
\partial^\nu)A_\mu^a
+\frac{1}{2} (\partial_5 A_\nu^a-\partial_\nu A_5^a)^2 \right. \nonumber \\
&& \left. -
\frac{1}{2\xi} \left( \partial_\mu A^{a\mu}
-\xi (\partial_5 A_5^a+g F^a_i \chi_i)\right)^2
+\frac{1}{2} \partial_\mu \chi_i \partial^\mu \chi_i  +\frac{1}{2}
g^2 F^a_iF^b_i A_\mu^a A^{\nu b} +\partial_\mu \chi_i A^{\mu a} g F^a_i
\right. \nonumber \\ && \left. -
\frac{1}{2} (\partial_5 \chi_i+g A^a_5 F^a_i)^2 -M^2_{ij} \chi_i\chi_j\right).
\end{eqnarray}
Note, that we have lowered all 5 indices.
Here $F^a_i = T^a_{ij} \langle \Phi_j \rangle$, which as always is
non-vanishing only in the directions that would correspond to the Goldstone
modes of the scalar potential.
Note, that the physics here is quite different than in the case
with no bulk scalars. Before the $A_5^a$'s were the would-be Goldstone modes
eaten by the massive gauge fields. Now this will change, and there is an
explicit
mass term for one combination of the $A_5$'s and the Goldstone components
of the $\chi$'s: $\partial_5 \chi_i+g A^a_5 F^a_i$. These fields will be
physical modes, that do not decouple even in the unitary gauge
$\xi \to \infty$. The other combination $\partial_5 A_5^a+g F^a_i \chi_i$
will provide the longitudinal modes of the gauge boson KK towers and will
disappear in the unitary gauge. Varying this action we get the linearized
bulk equations of motion
\begin{eqnarray}
&&
\partial_\sigma \partial^\sigma A^{a\,\nu}
-\partial_5^2 A^{a\, \nu}
-\left(1-\sfrac{1}{\xi}\right) \partial^\nu \partial_\sigma A^{a\, \sigma}
+ g^2 F^a_iF^b_iA^{\nu b}=0,
\nonumber \\
&&
\vphantom{\sfrac{1}{\xi}} \partial_\sigma \partial^\sigma A^{a}_5
- \xi \partial_5^2 A^{a}_5
-(\xi -1)g (\partial_5 \chi_i) F^a_i
-\xi g \chi_i \partial_5 F^a_i
+ g^2 F^a_iF^b_iA_5^b
=0\ ,
\nonumber \\
&&
\partial_\sigma \partial^\sigma \chi_i
-\partial_5^2 \chi_i
+(\xi-1)g (\partial_5 A_5^a) F^a_i
-g A_5^a \partial_5 F^a_i
+ \xi g^2 F^a_i F^a_j \chi_j
+ M^2_{ij}\chi_j
=0
\ .
\end{eqnarray}
The BC's will be modified to
\begin{eqnarray}
&&  F^{a}_{\nu 5} \, \delta A^{a \nu}{}_{|0,\pi R} =0,
\\
        \label{bcHiggs}
&&  (\partial_\sigma A^{a\sigma} - \xi \partial_5 A^a_5 -
\xi g \chi_i F^a_i) \delta A^a_5{}_{|0,\pi R} =0,
\\
&&
(\partial_5 \chi_i + g A_5^a F^a_i) \delta \chi_i=0.
\end{eqnarray}
A consistent set of BC's is obtained by taking the previous
set of BC from (\ref{bc}) and add the condition $\chi_i=0$ on the endpoints.
Note, that this does not imply the Higgs VEV on the brane has to vanish, since
the $\chi$'s are the fluctuations around the expectation value.

\subsection{Gauge theory with a scalar localized at the endpoint}
\label{app:braneHiggs}

Finally let us consider the case when one has a boundary scalar field
$\phi$, at $y=0$. The Lagrangian will be modified to
\begin{eqnarray}
        \label{eq:lagrhiggsbrane1}
&& \mathcal{S}=
\int d^4x\,\int_0^{\pi R} dy \left(
-\frac{1}{4} F_{\mu\nu}^a F^{a\mu\nu} -\frac{1}{2} F_{5\nu}^a F^{a5\nu}-
\frac{1}{2\xi} (\partial_\mu A^{a\mu} -\xi \partial_5 A_5^a)^2 \right)
\nonumber \\
&& +\int d^4x \left(\frac{1}{2} D_\mu \phi_i D^\mu \phi_i -V(\phi )
-\frac{1}{2\xi} ({\partial_\mu A^{\mu a}}_{|_0} -\xi g F^a_i \chi_i)^2\right).
\end{eqnarray}
Expanding to quadratic order we get that the action is
\begin{eqnarray}
        \label{eq:lagrhiggsbrane}
&& \mathcal{S}=
\int d^4x\,\int_0^{\pi R} dy \ {\cal L}_{bulk}
+\int d^4x \left( \frac{1}{2} \partial_\mu \chi_i \partial^\mu \chi_i
+ g^2 F^a_i F^b_i {A_\mu^a}_{|_0} {A^{\mu b}}_{|_0} +
g \partial^\mu \chi_i F^a_{i} {A^a_\mu}_{|_0} \right. \nonumber \\ &&
\left. -\frac{1}{2\xi} ({\partial_\mu A^{\mu a}}_{|_0} -\xi g F^a_i \chi_i)^2
-\frac{1}{2} M^2_{ij}\chi_i \chi_j \right)\,
\end{eqnarray}
where we now had to add a gauge fixing term both in the bulk and on the
brane. The bulk equations of motions will be as in (\ref{bulkeom}),
the BC's at $y=\pi R$ will be the ones given in (\ref{bc}), while
the BC's at $y=0$ will now be given by
\begin{eqnarray}
&&  (F^{a}_{\nu 5} +g^2 F^a_i F^b_i A_\nu^b+\frac{1}{\xi}\partial_\nu
\partial_\mu A^{a\mu})\, \delta A^{a \nu}{}_{|0} =0,
\\
        \label{bcbrane}
&&  (\partial_\sigma A^{a\sigma} - \xi \partial_5 A^a_5)
\delta A^a_5{}_{|0} =0,  \\
&& (-\partial_\mu \partial^\mu \chi_i -\xi g^2 F^a_i F^a_j \chi_j -M^2_{ij})
\delta \chi_i
=0.
\end{eqnarray}
In the limit $\xi \to \infty$ we get the usual unitary gauge where both
$\chi_i$'s and $A_5$'s (assuming there are no $A_5$ zero modes) are
decoupled, and one is left with the physical KK tower of $A_\mu$ and the
non-Goldstone scalar modes (the physical Higgses) which are orthogonal to the
directions $F^a_i\chi_i$. In this limit the BC for
the gauge fields will be of the form
\begin{equation}
\partial_y A_\mu^a{}_{|0,\pi R}= V_{0,\pi R}^{ab} A^b_\mu {}_{|0,\pi R}.
\label{mixedbcagain}
\end{equation}
We will refer to these mass term induced BC's as mixed BC's. Note that
these are mixed BC's still ensure the hermiticity (self-adjointness)
of the Hamiltonian. These are the BC's that should be used
for the KK expansion of the gauge fields.

\section{Sum Rules and Unitarity in Deconstruction}
\label{app:dec}
\setcounter{equation}{0}
\setcounter{footnote}{0}

 It was suggested~\cite{deconstruction} two years ago  that the physics of extra dimensions can be recovered
in the infrared in terms of a product of 4D gauge groups connected to each others by
some link fields (for the deconstruction of
supersymmetric theories see~\cite{susydec}).
We would like to see in this appendix how this correspondence
operates as far as the high energy behavior for the amplitude of elastic scattering
of massive gauge bosons is concerned (see also~\cite{Sekhar2} for similar computations).

Generically speaking, we have a set of gauge fields, $A^i_\mu (i=1,\ldots ,N)$, living
on the sites of a lattice. For simplicity, we will assume that all the gauge
group have the same gauge coupling $g$. The dynamics of the link fields leads
to a breaking of the product gauge group and, accordingly, (some of) the gauge bosons
acquire a mass (see~\cite{orbifolddec} for some phenomenological models mimicking
extra dimensional orbifold models).
The mass eigenstates are linear combinations of the site gauge fields
\begin{equation}
A_\mu^{(m)} = \sum_{i=1}^N \alpha^{(m)}_i A_\mu^i
\end{equation}
where the coefficients $\alpha^{(m)}_i$ define an orthonormal basis
\begin{equation}
\sum_{i=1}^{N} \alpha^{(m)}_i \alpha^{(n)}_i = \delta^{mn},
\ \
\sum_{m=1}^{N} \alpha^{(m)}_i \alpha^{(m)}_j = \delta_{ij}.
\end{equation}
The cubic and quartic couplings~(\ref{eq:cubic})-(\ref{eq:quartic}) are replaced by
\begin{eqnarray}
g_{cubic} & \to & g_{mnk}=g \sum_{i=1}^N   \alpha^{(m)}_i  \alpha^{(n)}_i  \alpha^{(k)}_i ,
\\
g_{quartic}^2 & \to & g_{mnkl}= g \sum_{i=1}^N \alpha^{(m)}_i  \alpha^{(n)}_i
\alpha^{(k)}_i \alpha^{(l)}_i.
\end{eqnarray}
The expansion with the energy of the elastic scattering amplitude will be of the usual form:
\begin{equation}
\mathcal{A}= A^{(4)} \frac{E^4}{M_n^4} +
A^{(2)} \frac{E^2}{M_n^2}+ A^{(0)}.
\end{equation}
The term growing with $E^4$ is proportional to
\begin{equation}
A^{(4)} \propto g^2
\left(
\sum_{i=1}^{N}  \alpha^{(m)\, 4}_i
-\sum_{k=1}^{N} \sum_{i,j=1}^{N} \alpha^{(m)\, 2}_i \alpha^{(m)\, 2}_j \alpha^{(k)}_i \alpha^{(k)}_j
\right).
\end{equation}
Again this term is just cancelling because of the orthonormality of the eigenstates.

The expression for the amplitude that grows with $E^2$, after using the orthonormality relation, is found to
be proportional to
\begin{eqnarray}
A^{(2)} \propto
g^2
\left(
4   M_{n}^2 \sum_{i=1}^{N}  \alpha^{(m)\, 4}_i
-3 \sum_k^N   M_{k}^2  \sum_{i,j=1}^N  \alpha^{(m)\, 2}_i \alpha^{(m)\, 2}_j \alpha^{(k)}_i \alpha^{(k)}_j
\right).
\end{eqnarray}
Unlike in the extra dimensional case and due to the absence of 5D Lorentz invariance,
there is no generic expression for the sum rule. In general the sum will not cancel but rather
will be suppressed by power of the replication number, $N$. There is one simplification which allows
to perform the sum over the mass eigenstates: indeed, from the definition of the eingenstates, we get
\begin{equation}
 \sum_k^N   M_{k}^2  \sum_{i,j=1}^N  \alpha^{(m)\, 2}_i \alpha^{(m)\, 2}_j \alpha^{(k)}_i \alpha^{(k)}_j
= \sum_{i,j=1}^N   \alpha^{(m)\, 2}_i \alpha^{(m)\, 2}_j M^2_{ij}
\end{equation}
where $M^2_{ij}$ is the square mass matrix in the theory space. Therefore, in order to evaluate
the elastic scattering amplitude, we do not need to fully diagonalize the mass matrix and to find
all the eigenvectors: the computation of the elastic scattering amplitude of a particular
mass eigenstate requires only the knowledge of the decomposition of this particular
eigenstate in terms of the theory space gauge bosons.

We will evaluate the scattering amplitude in two explicit examples.
Let us first consider the deconstruction version of the $SU(2)\to U(1)$ orbifold breaking
of Section~\ref{sec:orbifold}. The matter content of the model is the following
\begin{equation*}
\begin{array}{c|cccccc}
& SU(2)_1 & SU(2)_2 &  \multicolumn{2}{c}{\hdots} & SU(2)_{N-1} & U(1) \\
\hline
\phi_1 & \Yfund & \overline{\Yfund} \\
\phi_2 & & \Yfund & \overline{\Yfund} \\
\vdots  & & & \multicolumn{2}{c}{\ddots} \\
\phi_{N-2} & & & & \Yfund & \overline{\Yfund} \\
\phi_{N-1} & & & & & \Yfund & 1/2
\end{array}
\end{equation*}
The breaking to a single $U(1)$ is achieved by giving VEVs to the link fields $\phi_{i}, i=1\ldots N-1$,
\begin{equation}
\langle \phi_i \rangle = \frac{v}{\sqrt{2}}\, \mathbf{1} \ {\rm for} \ i=1\ldots N-2, \quad
\langle \phi_{N-1}\rangle = \frac{v}{\sqrt{2}} \left( \begin{array}{c} 0 \\ 1 \end{array} \right).
\end{equation}
The spectrum contains a massless photon, $\gamma^{(0)}$, and its KK excitations,  $\gamma^{(k)}$
$k=1\ldots N-1$, as well as a finite tower of massive charged gauge bosons,
$W^{\pm\, (k)}$ $k=1\ldots N-1$. In the lattice site basis, the $N\times N$ photon mass matrix has the form
\begin{equation}
\frac{g^2 v^2}{4}
\left(
\begin{array}{ccccc}
1 & -1 \\
-1 & 2 & -1 \\
& & \ddots \\
& & -1 & 2 & -1 \\
& & & -1 & 1
\end{array}
\right)
\end{equation}
which is diagonalized by  ($k=0 \ldots N$)
\begin{equation}
\gamma^{(k)} = \sqrt{\frac{2}{2^{\delta_k}N}} \sum_{i=1}^N \cos \frac{(2i-1)k\pi}{2N} \gamma^{i},
\quad
M^{\gamma}_k = gv \sin \frac{k \pi}{2N}.
\end{equation}
The  $(N-1)\times (N-1)$ $W$ mass matrix is
\begin{equation}
\frac{g^2 v^2}{4}
\left(
\begin{array}{ccccc}
1 & -1 \\
-1 & 2 & -1 \\
& & \ddots \\
& & -1 & 2 & -1 \\
& & & -1 & 2
\end{array}
\right)
\end{equation}
which is diagonalized by   ($k=1\ldots N-1$)
\begin{equation}
W^{(k)} = \sqrt{\frac{4}{2N-1}}  \sum_{i=1}^{N-1} \cos \frac{(2i-1)(2k-1)\pi}{4N-2} W^{i},
\quad
M^{W}_k = gv \sin \frac{(2k-1) \pi}{4N-2}.
\end{equation}
This decomposition allows to evaluate as a function of $N$
the sum rule appearing in the elastic scattering amplitude. For instance for
the $W^{(1)}  W^{(1)} \to  W^{(1)}  W^{(1)}$ scattering,we get
\begin{equation}
4   \sum_{i=1}^{N-1}  \alpha^{(1)\, 4}_i
-3 \sum_{i,j=1}^{N-1}  \alpha^{(1)\, 2}_i \alpha^{(1)\, 2}_j
\frac{M_{ij}^{2}}{M_1^{W\, 2}}
\sim 3.70\, N^{-3}.
\end{equation}

We will not discuss in details the deconstruction version of the left-right model
of Section~\ref{sec:LR} but we would like to present how to deconstruct
the outer-automorphism like BC's. To this end, let us simply consider
a 5D $SU(2)_L\times SU(2)_R$ model broken to $SU(2)_D$ by
the exchange of the two $SU(2)$'s at one end-point of the interval.
The matter content of the deconstructed version of the model is the following
\begin{equation*}
\begin{array}{c|ccccccccc}
& SU(2)^{L}_N & SU(2)^{L}_{N-1} & \hdots   & SU(2)^{L}_1
& SU(2)_D
&  SU(2)^{R}_1  & \hdots & SU(2)^{R}_{N-1}  & SU(2)^{R}_N
\\
\hline
\phi_{N-1}^L & \overline{\Yfund} & \Yfund \\
\phi_{N-2}^L & & \overline{\Yfund}   \\
\vdots    & & & \ddots \\
\phi_1^L & & & & \Yfund \\
\phi & & & & \overline{\Yfund} & \Ysymm & \Yfund \\
\phi_1^R & & & & & & \overline{\Yfund} \\
\vdots   & &  & & & & & \ddots    \\
\phi_{N-2}^R & & & & & & & & \Yfund \\
\phi_{N-1}^R & & & & & & & & \overline{\Yfund} & \Yfund
\end{array}
\end{equation*}
Note the presence of a link field, $\phi$, charged under three gauge groups.
The breaking to a single $SU(2)$ is achieved through the VEV pattern
\begin{equation}
\langle \phi^{L,R}_i \rangle = \frac{v}{\sqrt{2}} \mathbf{1}, i=1\ldots N-1,
\quad
\langle \phi^\alpha{}_\beta{}^\gamma{}_\delta \rangle =
\frac{v}{\sqrt{2}}
\left(\delta^\alpha{}_\beta \, \delta^\gamma{}_\delta -
\frac{1}{2} \delta^\gamma{}_\beta\, \delta^\alpha{}_\delta \right).
\end{equation}
Here $\alpha$ is an $SU(2)_1^L$ index, $\delta$ an $SU(2)_1^R$ index,
while $\beta$ and $\gamma$ are $SU(2)_D$ indices.
Reproducing the KK towers of the two particular linear combinations
$(A^L\pm A^R)/\sqrt{2}$ in the 5D model,  there are actually two kinds of gauge boson eigenvectors:
\begin{eqnarray}
{\rm ``+"\  states:} && (x_N, x_{N-1}, \ldots, x_1, y, x_1, \ldots, x_{N-1}, x_N),\\
{\rm ``-" \ states:} && (x_N, x_{N-1}, \ldots, x_1,  0,-x_1, \ldots, -x_{N-1}, -x_N).
\end{eqnarray}
For the ``+" eigenstates, we have relied on a numerical
diagonalization. The ``$-$" eigenstates, however,  can be found analytically from the zeros of
the Chebyshev polynomial of order $2N+1$, which allows for an analytical
estimation of the $E^2$ terms in the elastic scattering of the ``$-$" eigenstates.
For instance, for the scattering of the lightest massive states, we found a sum rule
that, again, scales like $1/N^3$
\begin{equation}
4   \sum_{i=1}^{2N+1}  \alpha^{(1^-)\, 4}_i
-3 \sum_{i,j=1}^{2N+1}  \alpha^{(1^-)\, 2}_i \alpha^{(1^-)\, 2}_j
\frac{M_{ij}^{2}}{M_{1^-}^{2}}
\sim 1.84\, N^{-3}.
\end{equation}
%


\end{document}